\documentclass[10pt,journal,compsoc]{IEEEtran}

\usepackage{subfigure}
\usepackage{multirow}
\usepackage{algorithm}  
\usepackage{algorithmicx}  
\usepackage{algpseudocode}  
\usepackage{amsmath}  
\usepackage{enumitem}
\usepackage{wrapfig}
\usepackage{cite}
\usepackage{amsmath,amssymb,amsfonts}
\usepackage{graphicx}
\usepackage{url}
\usepackage{booktabs} 
\usepackage[normalem]{ulem}
\usepackage[T1]{fontenc}

\newcommand{\ie}{\emph{i.e., }}
\newcommand{\eg}{\emph{e.g., }}

\newcommand{\wrt}{\emph{w.r.t. }}
\newcommand{\cf}{\emph{cf. }}
\newcommand{\aka}{\emph{a.k.a. }}

\floatname{algorithm}{Algorithm}

\begin{document}

\title{Learning Robust Recommender from Noisy \\Implicit Feedback}

\author{Wenjie~Wang,
        Fuli~Feng,
        Xiangnan~He,
        Liqiang~Nie,
        Tat-Seng~Chua
\IEEEcompsocitemizethanks{
\IEEEcompsocthanksitem This is an extension of our previous work~\cite{wang2021denoising} (WSDM 2021). We substantially extend it from the following aspects: 1) we propose to utilize sparse extra feedback (\eg ratings) to help denoising recommendation; 2) we develop two training strategies to incorporate extra feedback into ADT: finetuning and warm-up training. Besides, we propose one new causality-based inference strategy --- colliding inference; 3) we conduct extensive experiments and analysis on the three strategies in Section \ref{sec:experiment}; and 4) we add related work about causal inference and robust recommendation in Section \ref{sec:related_work}.
\IEEEcompsocthanksitem W. Wang, F. Feng and TS. Chua are with School of Computing, National University of Singapore, Computing 1, Computing Drive, 117417, Singapore.
E-mail: \{wenjiewang96, fulifeng93\}@gmail.com, dcscts@nus.edu.sg. Corresponding author: Fuli Feng.
\protect
\IEEEcompsocthanksitem X. He is with School of Information Science and Technology, University of Science and Technology of China, Hefei, China.
E-mail: xiangnanhe@gmail.com.
\protect
\IEEEcompsocthanksitem L. Nie is with the School of Computer Science and Technology, Shandong University, Qingdao, China. E-mail: nieliqiang@gmail.com.
}
}


\IEEEtitleabstractindextext{%
\begin{abstract}
The ubiquity of implicit feedback makes it indispensable for building recommender systems. However, it does not actually reflect the actual satisfaction of users. For example, in E-commerce, a large portion of clicks do not translate to purchases, and many purchases end up with negative reviews. As such, it is of importance to account for the inevitable noises in implicit feedback. However, little work on recommendation has taken the noisy nature of implicit feedback into consideration. 
In this work, we explore the central theme of denoising implicit feedback for recommender learning, including training and inference. 
By observing the process of normal recommender training, we find that noisy feedback typically has large loss values in the early stages. Inspired by this observation, we propose a new training strategy named \textit{Adaptive Denoising Training} (ADT), which adaptively prunes the noisy interactions by two paradigms (\ie \textit{Truncated Loss} and \textit{Reweighted Loss}). Furthermore, 
we consider extra feedback (\eg rating) as auxiliary signal and propose three strategies to incorporate extra feedback into ADT: \textit{finetuning}, \textit{warm-up training}, and \textit{colliding inference}. 
We instantiate the two paradigms on the widely used binary cross-entropy loss and test them on three representative recommender models. Extensive experiments on three benchmarks demonstrate that ADT significantly improves the quality of recommendation over normal training without using extra feedback. Besides, the proposed three strategies for using extra feedback largely enhance the denoising ability of ADT. 
\end{abstract}

\begin{IEEEkeywords}
Recommendation, Denoising Implicit Feedback, Adaptive Denoising Training, Colliding Inference
\end{IEEEkeywords}}

\maketitle
\IEEEdisplaynontitleabstractindextext
\IEEEpeerreviewmaketitle

\section{Introduction}\label{sec:intro}

\IEEEPARstart{R}{ecommender} systems have been widely used to mine user preference over various items in many online services such as news portals~\cite{Lu2018Between} and social networks~\cite{He2017Neural}. As the clue to user choices, implicit feedback (\eg click and purchase) are typically the default choice to train a recommender model due to their large volume. However, prior work~\cite{Hu2008CF, Lu2018Between, Wen2019Leveraging} points out the gap between implicit feedback and the actual user satisfaction due to the existence of \textit{noisy interactions} (\aka \textit{false-positive interactions}) where the users dislike the interacted items. For instance, in E-commerce, a large portion of purchases end up with negative reviews or being returned. This is because implicit interactions are easily affected by the first impression of users and factors \cite{wang2020click} like caption bias \cite{Hofmann2012on} and position bias \cite{Jagerman2019To}. Moreover, existing studies~\cite{ Wen2019Leveraging} have validated the detrimental effect of such 
noisy interactions on user experience. However, little work on recommendation has considered the noisy nature of implicit feedback.

In this work, we argue that false-positive interactions would prevent a recommender model from learning the actual user preference, leading to low-quality recommendations.
Table~\ref{table1} provides empirical evidence on the negative effects of false-positive interactions when we train a competitive recommender model, Neural Matrix Factorization (NeuMF)~\cite{He2017Neural}, on two real-world datasets. 
We construct a ``clean'' testing set by removing the false-positive interactions for recommender evaluation\footnote{Each false-positive interaction is identified by some reliable feedback, \eg rating score ([1, 5]) $ < 3$, indicating that the user is dissatisfied with the item. Refer to Section \ref{sec:study} for more details.}. As shown in Table \ref{table1}, training NeuMF with false-positive interactions (\ie \textit{normal training}) results in sharp performance drop as compared to the NeuMF trained without false-positive interactions (\ie \textit{clean training}).
Therefore, it is essential to denoise implicit feedback for recommender learning.

\begin{figure}[t]
\setlength{\abovecaptionskip}{0cm}
\includegraphics[scale=0.52]{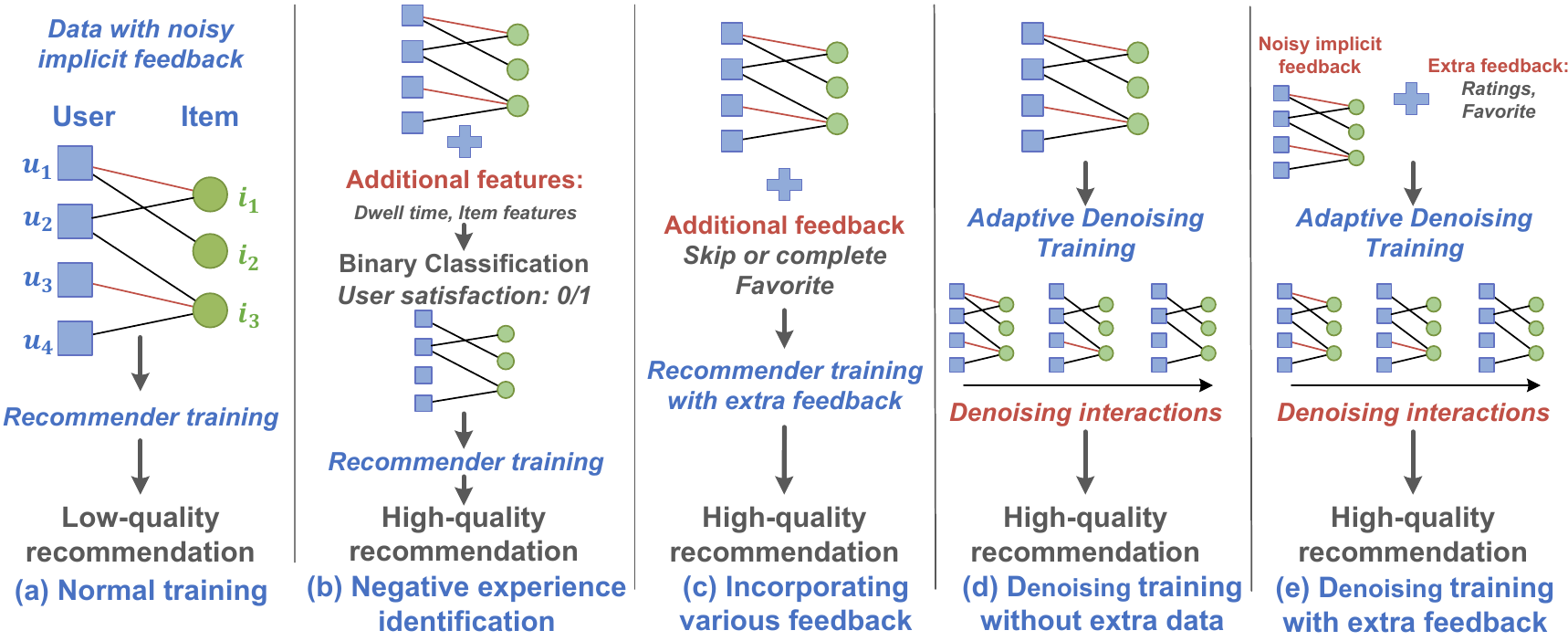}
\caption{The comparison between normal training (a); two prior solutions (b) and (c); and the proposed denoising training without additional data (d) and with extra feedback (e). Note that red lines in the user-item graph denote false-positive interactions, and extra feedback usually cannot identify all false-positive ones due to the sparsity issue.}
\label{Figure1}
\vspace{-0.2cm}
\end{figure}

Existing studies \cite{fox2005Evaluating, Kim2014Modeling, Yang2012Exploiting} on denoising implicit feedback mainly fall into two categories: 1) negative experience identification~\cite{Kim2014Modeling}; and 2) the incorporation of various feedback \cite{Yang2012Exploiting, Yi2014Beyond}. As illustrated in Figure~\ref{Figure1}(b), the former identifies the false-positive interactions with additional user behaviors (\eg dwell time and gaze pattern) and auxiliary item features (\eg length of the item description)~\cite{Lu2018Between} before recommender training. Besides, Figure \ref{Figure1}(c) shows that the latter leverages extra feedback (\eg favorite and skip) to prune the effects of false-positive interactions~\cite{Yi2014Beyond}, for example, multi-task learning~\cite{Gao2019LearningTR}. A key limitation with these methods is that they heavily rely on extra feedback or auxiliary item features, which might be hard to collect. Furthermore, extra feedback (\eg rating or favorite) is often of a smaller scale, which will suffer from the sparsity issue. For instance, many users do not give any extra feedback after watching a movie or purchasing a product~\cite{Hu2008CF}.

%
%
%


This work aims to automatically reduce the negative effect of false-positive interactions, where we mainly count on the implicit interactions for denoising. 
Prior research on robust learning~\cite{lu2018MentorNet, han2018co} and curriculum learning~\cite{Bengio2009Curriculum} demonstrates that noisy samples are relatively harder to fit into models, exhibiting distinct patterns in the model training. We conduct preliminary experiments across different recommender models and datasets, obtaining a similar observation: \textit{the loss values of false-positive interactions are relatively larger in the early stages of training} (\cf Figure~\ref{fig:observation}). Consequently, false-positive interactions will mislead the recommender training in the early stages. Worse still, the recommender model ultimately fits the false-positive interactions due to its high representation capacity, which could be overfitting and hurt the generalization. In this light, a potential solution of denoising is to prune the interactions with large loss values in the early training stage.

Towards this end, we propose Adaptive Denoising Training (ADT), which dynamically prunes the large-loss interactions along the training process (illustrated in Figure \ref{Figure1}(d)). To avoid losing generality, we only modify the training loss, which can be applied to any differentiable models. In detail, we devise two paradigms to formulate the training loss: 1) \textit{Truncated Loss}, which discards the large-loss interactions dynamically with a threshold function, and 2) \textit{Reweighted Loss}, which adaptively assigns hard interactions (\ie large-loss ones) with smaller weights to reduce their influence on the recommender optimization.

To keep the fexibility of ADT in the scenarios with easily available extra feedback, we propose two training strategies to utilize extra feedback (Figure \ref{Figure1}(e)): 1) finetuning, which refines the recommender model with extra feedback after ADT; and 2) warm-up training, which trains the recommender model with extra feedback before ADT. 
To tackle the sparsity issue of extra feedback, we design an inference strategy, called colliding inference, which leverages the colliding effect~\cite{pearl2009causality} of extra feedback on the predictions of sparse users for better recommendation. From a causal view, extra feedback and the predictions of sparse users are $d$-separated due to the sparsity issue. By conditioning on a collider between them, we make them $d$-connected and leverage the colliding effect to augment the predictions of sparse users~\cite{Hu_2021_CVPR}.


We instantiate ADT on the widely used binary cross-entropy (CE) loss. On three benchmarks, we test the Truncated Loss and Reweighted Loss over three representative recommender models: Generalized Matrix Factorization (GMF)~\cite{He2017Neural}, NeuMF~\cite{He2017Neural}, and Collaborative Denoising Auto-Encoder (CDAE)~\cite{wu2016collaborative}. The results show significant performance improvements of ADT over normal training. Besides, we conduct extensive experiments to validate the effectiveness of the strategies that incorporate extra feedback into ADT. They further improve the performance of the vanilla ADT, and outperform some competitive baselines that use extra feedback for training. 
Codes and data are publicly available at: \url{https://github.com/WenjieWWJ/DenoisingRec}.

Our main contributions are summarized as:
\begin{itemize}[leftmargin=*]
    \item We reveal the negative effect of false-positive interactions and propose Adaptive Denoising Training to prune the large-loss interactions dynamically, which introduces two paradigms to formulate the training loss: Truncated Loss and Reweighted Loss.
    \item We develop three strategies (\ie finetuning, warm-up training, and colliding inference) to incorporate extra feedback into ADT, which provides the potential to further improve the denoising ability of ADT.
    \item Extensive experiments on three benchmarks validate the effectiveness of the proposed ADT and three strategies in improving the recommendation quality.
\end{itemize}


%
%
\begin{table}[t]
\setlength{\abovecaptionskip}{0cm}
\setlength{\belowcaptionskip}{0cm}
\caption{Results of the clean training and normal training over NeuMF. \#Drop denotes the relative performance drop of normal training as compared to clean training.}
\label{table1}
\centering
\resizebox{0.5\textwidth}{!}{
\begin{tabular}{l|cc|cc}
\hline
\textbf{Dataset}     & \multicolumn{2}{c|}{\textbf{Adressa}} & \multicolumn{2}{c}{\textbf{Amazon-book}} \\ 
\textbf{Metric}      & \textbf{Recall@20}    & \textbf{NDCG@20}   & \textbf{Recall@20}   & \textbf{NDCG@20}   \\ \hline \hline
\textbf{Clean training} & 0.4040                & 0.1963             & 0.0293               & 0.0159             \\ 
\textbf{Normal training}  & 0.3159                & 0.1886             & 0.0265               & 0.0145             \\ \hline
\textbf{\#Drop}      & 21.81\%               & 3.92\%            & 9.56\%               & 8.81\%             \\ \hline
\end{tabular}
}
\end{table}
%
%

\section{Study on False-Positive Feedback}\label{sec:study}

The influence of noisy training samples has been explored in many conventional machine learning tasks such as image classification~\cite{lu2018MentorNet, han2018co}. However, little work has considered denoising implicit feedback in recommendation, which is inherently different from conventional tasks due to the relations across training samples, \eg interactions on the same item. We investigate the effects of false-positive interactions on recommender training by comparing the performance under two settings: 1) normal training, which utilizes all observed user-item interactions, and 2) clean training, which discards false-positive interactions.
An interaction is identified as false-positive or true-positive one according to the explicit feedback. For instance, a purchase is false-positive if the following rating score ([1, 5]) < 3. Although such explicit feedback is sparse, the scale is sufficient to conduct a pilot experiment and construct a clean testing set. 
The recommender models are evaluated on a holdout clean testing set with only true-positive interactions kept, \ie the evaluation focuses on recommending more satisfying items to users. More details can be seen in Section~\ref{sec:experiment}.

\vspace{3pt}
\textbf{Results.}
From Table \ref{table1}, we can observe that, as compared to clean training, the performance of normal training drops significantly (\eg 21.81\% \wrt Recall@20 on Adressa). It validates the negative effects of false-positive interactions on recommending satisfying items to users. Worse still, if the recommender models under normal training are deployed, they will have higher risk to produce more false-positive interactions, which further hurt the user experience~\cite{Lu2018Between}. 

\section{Method}
\label{sec:method}

In this section, we detail the proposed ADT and three strategies to incorporate extra feedback into ADT. Prior to that, the task formulation of denoising recommender training and some observations that inspire the strategy design are introduced.

\subsection{\textbf{Task Formulation}}

Generally, the target of recommender training is to learn a scoring function $\hat{y}_{ui} = f(u, i | \Theta)$ to assess the preference of user $u$ over item $i$ with parameters $\Theta$. Ideally, $\Theta$ is learned from reliable feedback between users ($\mathcal{U}$) and items ($\mathcal{I}$) by minimizing a recommendation loss $\mathcal{L}$ (\eg the binary CE loss). Formally, $\Theta^* = \arg\min_{\Theta} \mathcal{L}({D}^{\ast})$ where ${D}^{\ast} = \{(u, i, y^{\ast}_{ui}) | u \in \mathcal{U}, i \in \mathcal{I}\}$, and $y^{\ast}_{ui} \in \{0, 1\}$ represents whether the user $u$ really prefers the item $i$. The recommender model with $\Theta^{\ast}$ would be reliable to generate high-quality recommendations. In practice, due to the lack of large-scale reliable feedback, recommender training is typically formalized as: $\bar{\Theta} = \arg\min_{\Theta} \mathcal{L}({\bar{D}})$, where ${\bar{D}} = \{(u, i, \bar{y}_{ui}) | u \in \mathcal{U}, i \in \mathcal{I}\}$ is a set of implicit interactions. $\bar{y}_{ui}$ denotes whether the user $u$ has interacted with the item $i$, such as click and purchase.

However, because noisy interactions in $\bar{D}$ would mislead the learning of $\Theta$, the typical recommender training might lead to a poor model (\ie $\bar{\Theta}$) that lacks generalization ability on the clean testing set. As such, we formulate a \textit{denoising recommender training} task as:
\begin{equation}\label{eqn:denoise}
\begin{aligned}
& {\Theta}^{\ast} = \arg\min_{\Theta} \mathcal{L}\big(denoise({\bar{D}})),
\end{aligned}
\end{equation}
aiming to learn a reliable recommender model with parameters $\Theta^{\ast}$ by denoising implicit feedback $\bar{D}$. Formally, by assuming the existence of inconsistency between $y^{\ast}_{ui}$  and $\bar{y}_{ui}$, we define noisy interactions (\aka false-positive interactions) as $\left\{ (u, i) | y^{\ast}_{ui}=0 \land \bar{y}_{ui}=1 \right\}$~\cite{wang2021denoising}. 

\begin{figure}[h]
\setlength{\abovecaptionskip}{0.cm}
\setlength{\belowcaptionskip}{-0.2cm}
  \centering 
  \hspace{-0.1in}
  \subfigure[Whole training process]{
    \label{fig2:subfig:a}
    \includegraphics[width=1.7in]{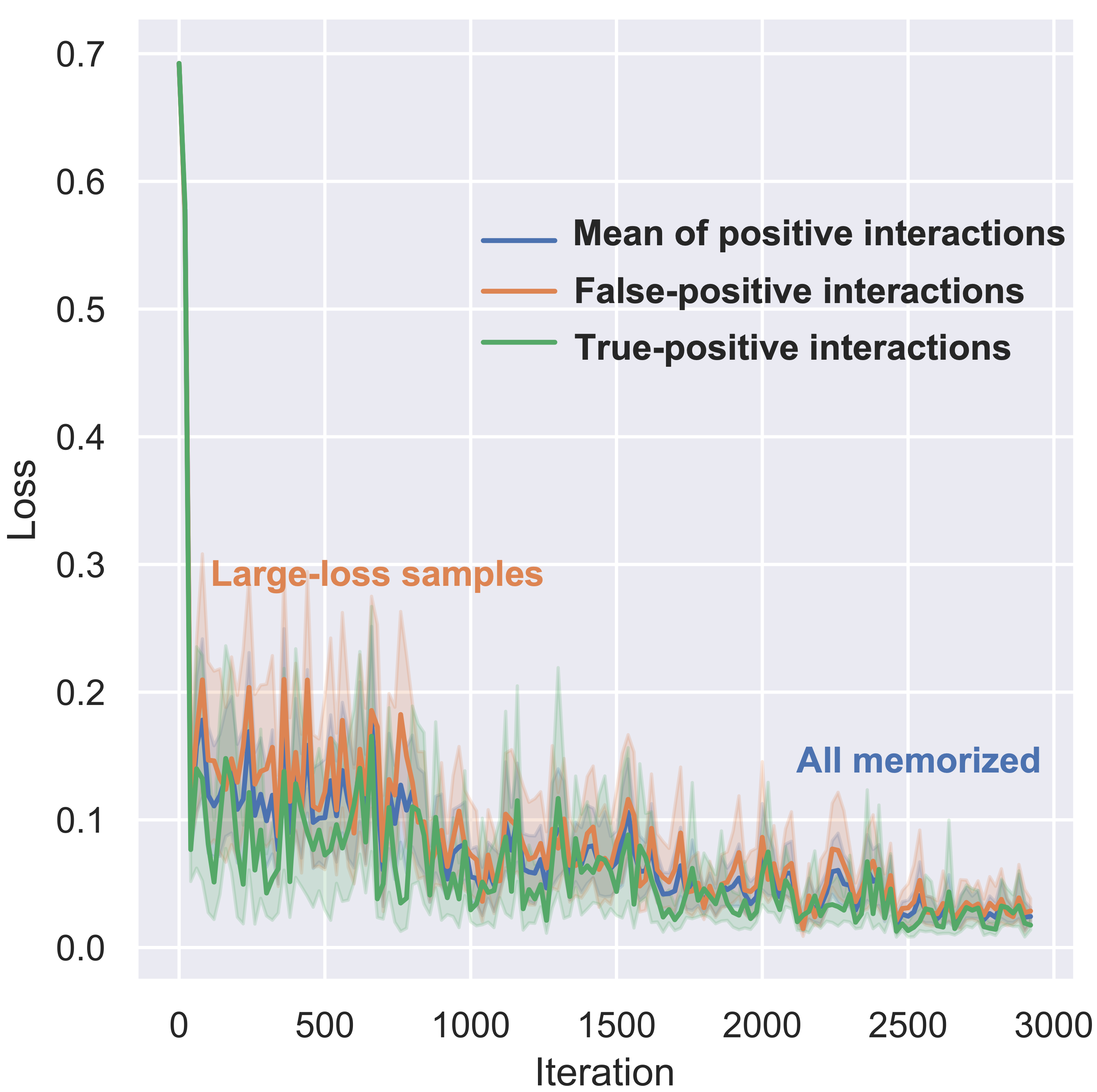}} 
  \subfigure[Early training stages]{ 
    \label{fig2:subfig:b} 
    \includegraphics[width=1.7in]{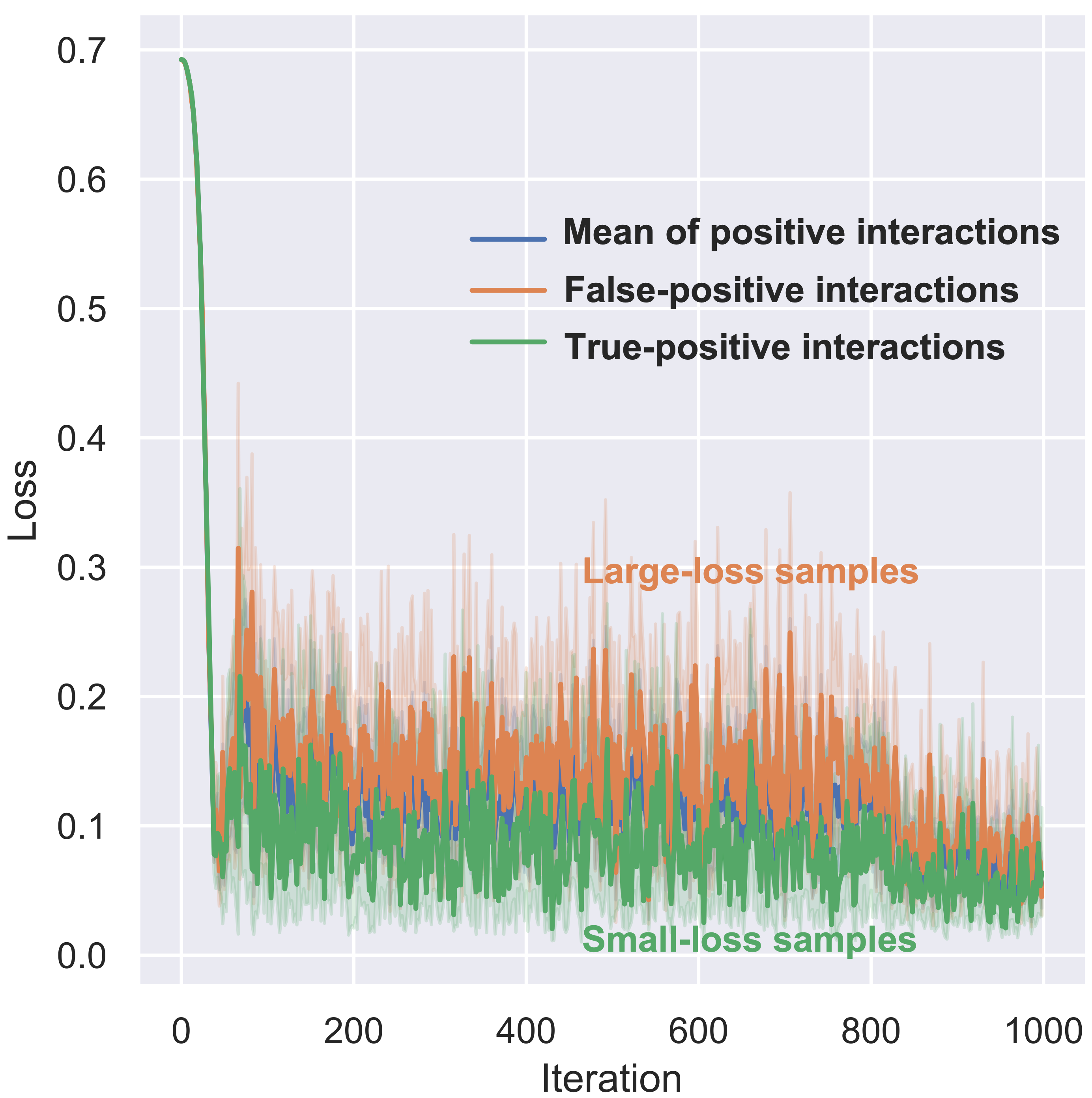}} 
  \caption{The training loss of true- and false-positive interactions on Adressa in the normal training of NeuMF. The false-positive interactions have larger loss values in the early training stage.} 
  \label{fig:observation}
\end{figure}
%
%

\subsection{Observations}\label{sec:observations}
\textbf{{False-positive interactions are harder to fit in the early stages.}}
According to the theory of robust learning~\cite{lu2018MentorNet, han2018co} and curriculum learning~\cite{Bengio2009Curriculum}, easy samples are more likely to be the clean ones and fitting the hard samples may hurt the generalization. To verify the existence of this phenomenon in recommendation, we compare the loss of true- and false-positive interactions along the process of normal training. Figure \ref{fig:observation} shows the results of NeuMF on the Adressa dataset. 
Similar trends are also found over other recommender models and datasets (see more details in Section \ref{sec:indepth}). 
From Figure \ref{fig:observation}, we observe that:
\begin{itemize}[leftmargin=*]
    \item Ultimately, the loss of both of true- and false-positive interactions converges to a stable state with close values, which implies that NeuMF fits both of them well. It reflects that deep recommender models can ``memorize'' all the training interactions, including the noisy ones. As such, if the data is noisy, the memorization will lead to poor generalization performance. 
    \item The loss values of true- and false-positive interactions decrease differently in the early stages of training. From Figure \ref{fig2:subfig:b}, we can see that the loss of false-positive interactions is clearly larger than that of the true-positive ones, which indicates that false-positive interactions are harder to memorize than the true-positive ones in the early stages. The reason might be that false-positive ones represent the items that users dislike, and they are more similar to the items the user didn't interact with. The findings also support the prior theory in robust learning and curriculum learning~\cite{han2018co, yu2019does, Bengio2009Curriculum}.
\end{itemize}
Overall, the results are also consistent with the memorization effect~\cite{arpit2017closer}: deep recommender models will learn the easy and clean patterns in the early stage, and eventually memorize all training interactions~\cite{han2018co} because of the strong representation ability.

\subsection{Adaptive Denoising Training}

According to the observations, we propose ADT strategies for recommender models, which estimate $P(y^{\ast}_{ui}=0 | \bar{y}_{ui}=1, u, i)$ according to the training loss. 
In particular, ADT either \textit{discards} or \textit{reweighs} the interactions with large loss values to reduce their influences on the recommender training. Towards this end, we devise two paradigms to formulate loss functions for denoising training without using extra feedback:
\begin{itemize}
\item \textit{Truncated Loss.} This is to truncate the loss values of hard interactions to 0 with a dynamic threshold function.
\item \textit{Reweighted Loss.} It adaptively assigns hard interactions with smaller weights during training.
\end{itemize}
Note that the two paradigms can be applied to various recommendation loss functions, \eg CE loss, square loss, and BPR loss~\cite{rendle2009bpr}. In the work, we take CE loss as an example to elaborate them.

%
%

\begin{figure}
\setlength{\abovecaptionskip}{0cm}
\setlength{\belowcaptionskip}{-0.5cm}
\centering
\includegraphics[scale=0.06]{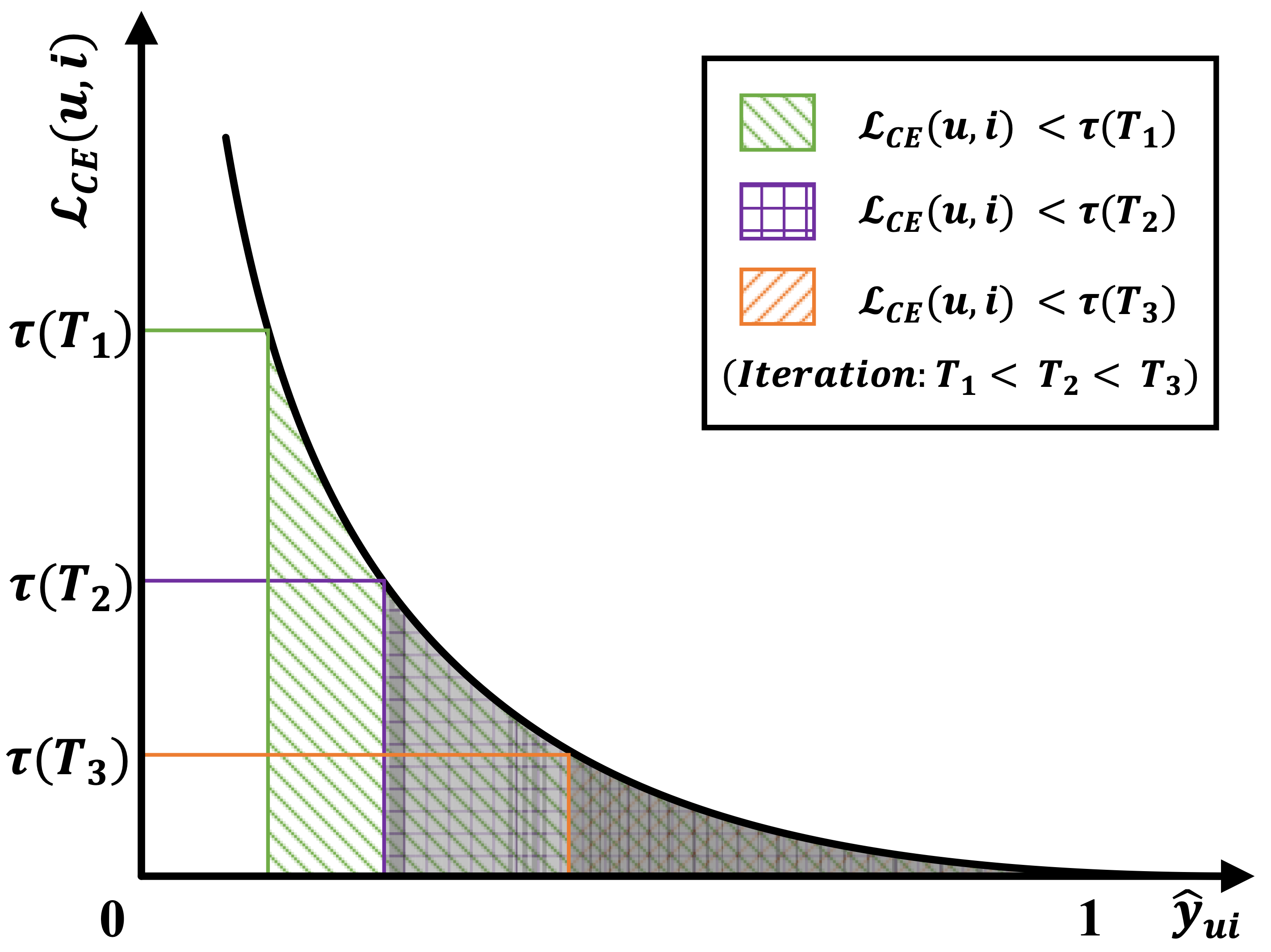}
\caption{Illustration of T-CE loss for the observed interactions (\ie $\bar{y}_{ui} = 1$). $T_i$ is the iteration number and $\tau(T_i)$ refers to the threshold. Dash area indicates the effective loss and the loss values larger than $\tau(T_i)$ are truncated during training.}
\label{fig:truncatedloss}
\end{figure}
%
%

\subsubsection{\textbf{Truncated Cross-Entropy Loss}}

Functionally speaking, the Truncated Cross-Entropy (shorted as T-CE) loss discards positive interactions according to the values of CE loss. Formally, we define it as:
%
%
\begin{equation}
\mathcal{L}_{\textit{T-CE}}(u, i) = 
\begin{cases}
0, & \mathcal{L}_{CE}(u, i) > \tau \land \bar{y}_{ui}=1 \\
\mathcal{L}_{CE}(u, i), & \text{otherwise,}
\end{cases}
\end{equation}
%
%
where $\tau$ is a pre-defined threshold. The T-CE loss removes any positive interactions with CE loss (\ie $\mathcal{L}_{CE}$) larger than $\tau$ from the training. While the T-CE loss is easy to interpret and implement, the fixed threshold may not work properly. This is because the loss value is decreasing with the increase of training iterations. Inspired by the dynamic adjustment ideas~\cite{kingma2014adam}, we replace the fixed threshold with a dynamic threshold function $\tau(T)$ \wrt the training iteration $T$, which changes the threshold value along the training process (Figure \ref{fig:truncatedloss}).
Besides, since the loss values vary across different datasets, it would be more flexible to devise $\tau(T)$ as a function of the drop rate $\epsilon(T)$. There is a bijection between the drop rate and the threshold, \ie we can calculate the threshold based on the drop rate $\epsilon(T)$ at each iteration.

Based on prior observations, a proper drop rate function should satisfy: 
1) $\epsilon(\cdot)$ should have an upper bound to limit the proportion of discarded interactions so as to prevent data missing;
2) $\epsilon(0) = 0$, \ie it should allow all the interactions to be fed into the models in the beginning; 
and 3) $\epsilon(\cdot)$ should increase smoothly from zero to its upper bound, so that the model can learn and distinguish the true- and false-positive interactions gradually.

To this end, we formulate the drop rate function as:
%
%
\begin{equation}
\epsilon(T) = min(\alpha T, \epsilon_{max}),
\end{equation}
%
%
where $\epsilon_{max}$ is an upper bound and $\alpha$ is a hyper-parameter to adjust the pace to reach the maximum drop rate. Here we increase the drop rate in a linear fashion rather than a more complex function such as a polynomial function or a logarithm function. Despite the expressiveness of these functions, they have more hyper-parameters, increasing cost of tuning a recommender model. The algorithm is formally explained in Algorithm \ref{algo:T-CE}. 

\begin{algorithm}[t]
    \footnotesize
	\caption{Adaptive Denoising Training with T-CE loss}  
	\label{algo:T-CE}
	\begin{algorithmic}[1]
		\Require trainable parameters $\Theta$, training interactions ${\bar{D}}$, the iterations $T_{max}$, learning rate $\eta$, $\epsilon_{max}$, $\alpha$, $\mathcal{L}_{CE}$, parameter optimization function $\nabla$.
		\For{$T = 1 \to T_{max}$} \algorithmiccomment{shuffle interactions every epoch}
		\State \textbf{Fetch} mini-batch data ${\bar{D}}_{pos}$ from ${\bar{D}}$
		\State \textbf{Sample} unobserved interactions ${\bar{D}}_{neg}$
		\State \textbf{Define} ${\bar{D}}_{T} = {\bar{D}}_{pos} \cup {\bar{D}}_{neg}$
		\State \textbf{Obtain} ${\hat{D}} = \mathop{\arg\max}\limits_{{\hat{D}}\subset {\bar{D}}_{pos}, |{\hat{D}}| = \lfloor \epsilon(T)*|{\bar{D}}_{\textit{T}}|\rfloor} \sum_{(u,i) \in {\hat{D}}} \mathcal{L}_{CE}(u, i | \Theta_{\textit{T-1}})$
		\State \textbf{Obtain} ${{D}^*} = {\bar{D}}_{T} - {\hat{D}}$
		\State \textbf{Update} $\Theta_{\textit{T}} = \nabla(\Theta_{\textit{T-1}}, \eta, \mathcal{L}_{CE}, {{D}^*})$
		\State \textbf{Update} $\epsilon(T) = min(\alpha T, \epsilon_{max})$
		\EndFor 
		\Ensure the optimized parameters $\Theta_{T_{max}}$ of the recommender model.
	\end{algorithmic}
\end{algorithm}

\subsubsection{\textbf{Reweighted Cross-Entropy Loss}}
Generally, the Reweighted Cross-Entropy (shorted as R-CE) loss down-weights the hard positive interactions. Formally,
%
%
\begin{equation}
\begin{aligned}
& \mathcal{L}_{\textit{R-CE}}(u, i) =  \omega(u, i) \mathcal{L}_{\textit{CE}}(u, i), 
\end{aligned}
\end{equation}
%
%
where $\omega(u, i)$ is a weight function that adjusts the contribution of an interaction to the training objective. To properly down-weight the hard interactions, the weight function $\omega(u, i)$ is expected to have the following properties:
1) it dynamically adjusts the weights of interactions during training;
2) the function makes the influence of a hard interaction weaker than an easy interaction;
and 3) the extent of weight reduction can be easily adjusted to fit different models and datasets.

\begin{figure}[t]
\setlength{\abovecaptionskip}{0.1cm}
\setlength{\belowcaptionskip}{-0.20cm}
\centering 
\hspace{-0.25in}
    \subfigure[R-CE loss for the observed positive interactions. The contributions of large-loss interactions are greatly reduced.]{
    \label{fig5:subfig:a}
    \includegraphics[width=1.6in]{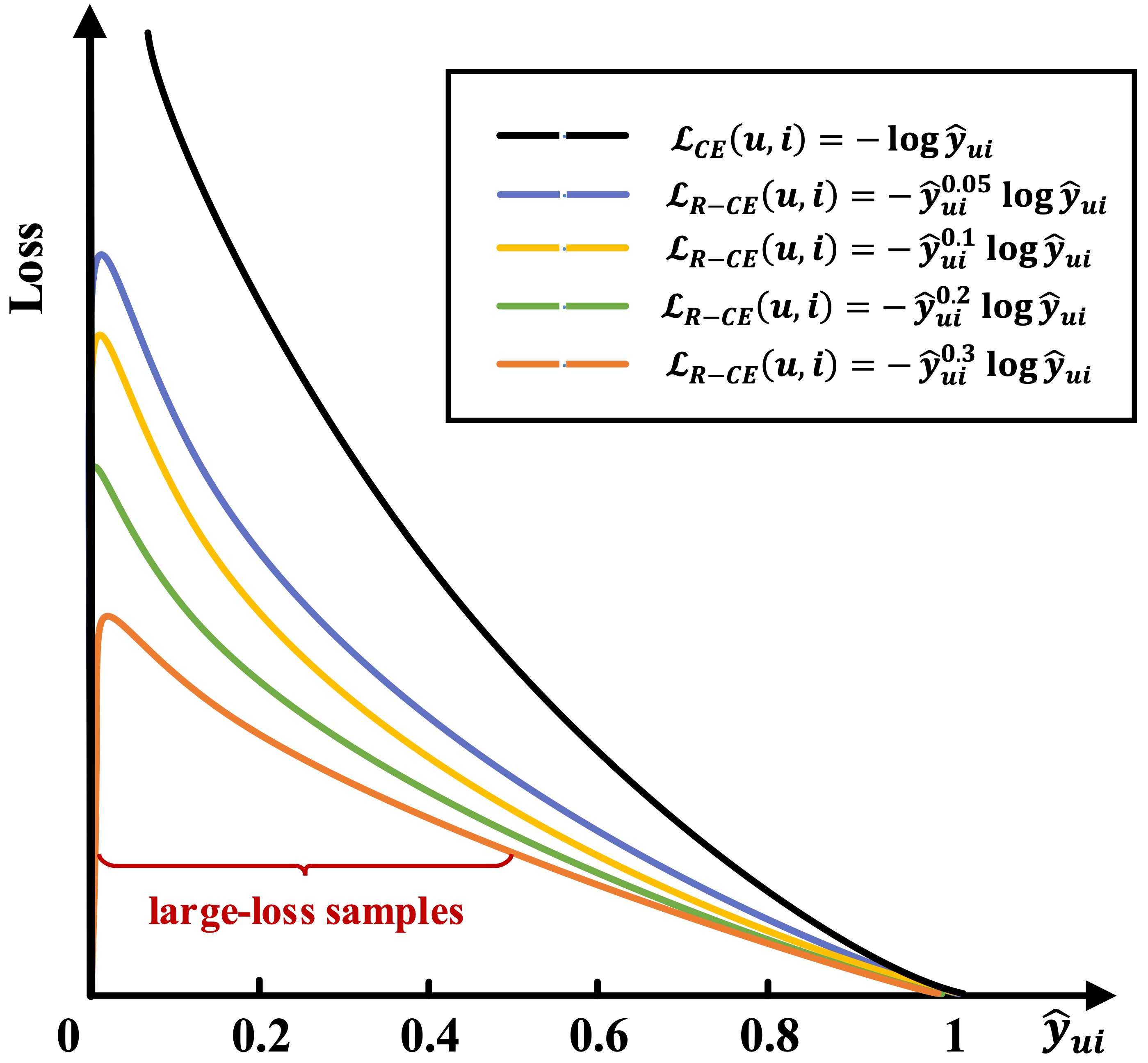}} 
    \hspace{0.1in}
    \subfigure[The weight function with different parameters $\beta$, where $\beta$ controls the weight difference between hard and easy interactions.]{
    \label{fig5:subfig:b} 
    \includegraphics[width=1.85in]{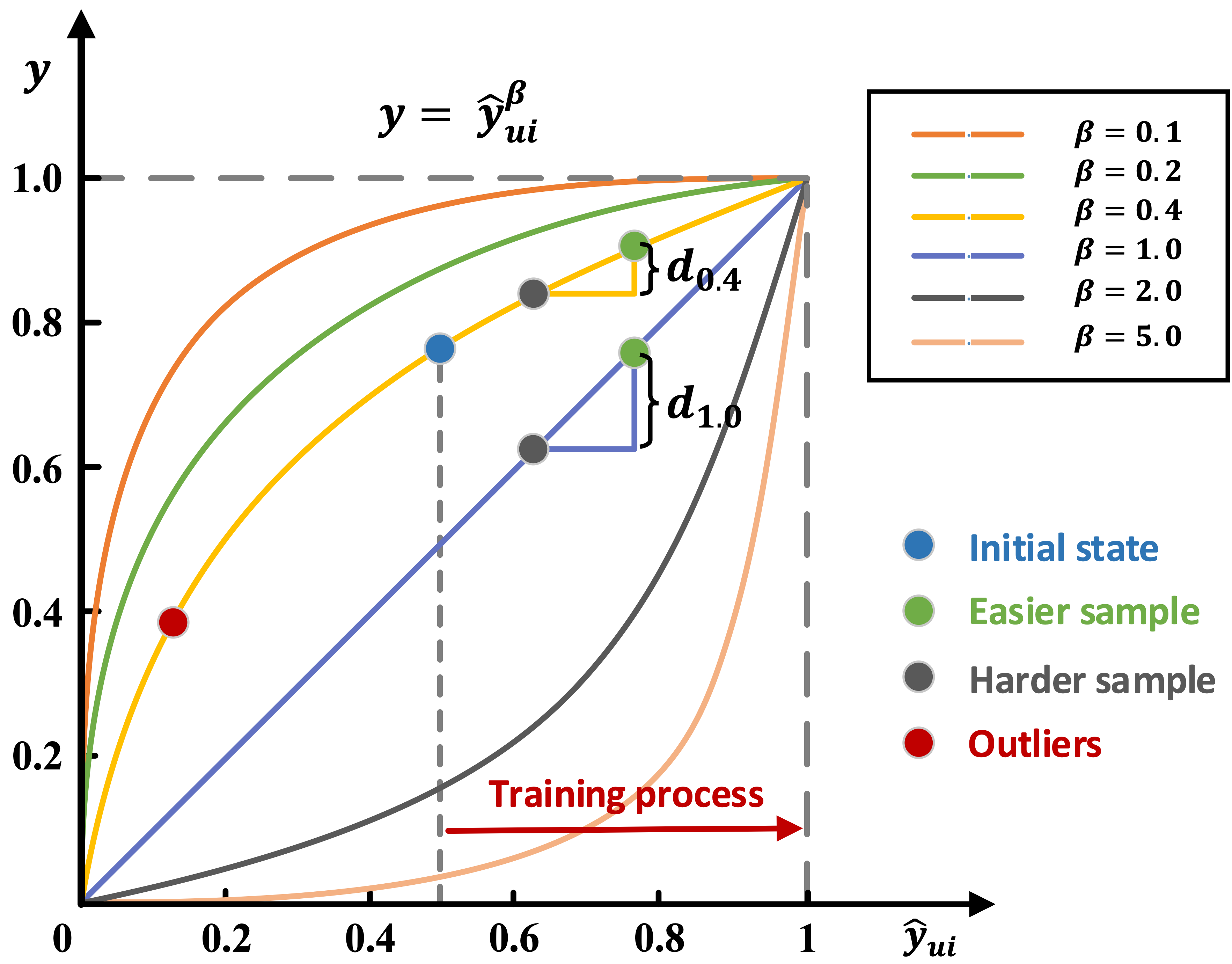}} 
\hspace{-0.25in}
\caption{Illustration and analysis of R-CE loss.} 
\label{fig:R-CE}
\end{figure}
%
%

Inspired by the success of Focal Loss~\cite{lin2017focal}, we estimate $\omega(u, i)$ according to the prediction score $\hat{y}_{ui}$. The prediction score is within $[0, 1]$ whereas the value of CE loss is in $[0, +\infty]$. And thus the prediction score is more accountable for further computation. Note that the smaller prediction score on the positive interaction leads to larger CE loss. 
Formally, we define the weight function as:
%
%
\begin{equation}
 \omega(u, i) = \hat{y}_{ui}^{\beta},
\end{equation}
%
%
where $\beta \in [0, +\infty]$ is a hyper-parameter to control the range of weights. 
From Figure \ref{fig5:subfig:a}, we can see that R-CE loss can significantly reduce the loss of hard interactions (\eg $\hat{y}_{ui} \ll 0.5$) as compared to the original CE loss. Furthermore, the proposed weight function satisfies the aforementioned requirements:

\begin{itemize}[leftmargin=*]
    \item It generates dynamic weights during training since $\hat{y}_{ui}$ changes along the training process.
    \item The interactions with extremely large CE loss (\eg the ``outlier'' in Figure \ref{fig5:subfig:b}) will be assigned with very small weights because $\hat{y}_{ui}$ is close to 0. Therefore, the influence of such large-loss interactions is largely reduced. 
    Besides, as shown in Figure \ref{fig5:subfig:b}, harder interactions always have smaller weights because the function $\hat{y}_{ui}^{\beta}$ monotonically increases \wrt $\hat{y}_{ui}$ when $\hat{y}_{ui} \in [0, 1]$ and $\beta \in [0, +\infty]$. As such, it can avoid that false-positive interactions dominate the optimization during training \cite{Yang2018robust}.
    \item The hyper-parameter $\beta$ dynamically controls the gap between the weights of hard and easy interactions. By observing the examples in Figure \ref{fig5:subfig:b}, we can find that: 1) the gap between the weights of easy and hard interactions becomes larger as $\beta$ increases from 0 to 1 (\eg $d_{0.4} < d_{1.0}$) when $\hat{y}_{ui} > 0.5$; and 2) the R-CE loss will degrade to a standard CE loss if $\beta$ is 0. 
\end{itemize}

Lastly, in order to prevent negative interactions with large loss values from dominating the optimization, we revise the weight function as:
%
%
\begin{equation}
\omega(u, i) = 
\begin{cases}
\hat{y}_{ui}^{\beta}, & \bar{y}_{ui}=1, \\
(1-\hat{y}_{ui})^{\beta}, & \text{otherwise.}
\end{cases}
\end{equation}
%
%
Indeed, it may provide a idea to alleviate the impact of false-negative interactions, which is left for future exploration.

\subsubsection{\textbf{In-depth Analysis}}

Due to depending totally on recommender models to identify false-positive interactions, the reliability of ADT needs to be discussed. Actually, existing work~\cite{lu2018MentorNet, han2018co} has pointed out the connection between the large loss and noisy interactions, and explained the underlying causality: the ``memorization'' effect of deep models~\cite{arpit2017closer}. 
We discuss the memorization effect of recommender models by experiments in Section \ref{sec:observations} and \ref{sec:indepth}. 

Another concern is that discarding hard interactions would limit the model's learning ability since some hard interactions may be more informative than the easy ones. Indeed, as discussed in the prior studies on curriculum learning~\cite{Bengio2009Curriculum}, hard interactions in the noisy data probably confuse the classifier rather than help it to establish the right decision surface. It's actually a trade-off between denoising and learning. In ADT, the $\epsilon(\cdot)$ of T-CE loss and $\beta$ of R-CE loss are used to control the balance.

\begin{figure}[t]
\setlength{\abovecaptionskip}{0cm}
\includegraphics[scale=0.6]{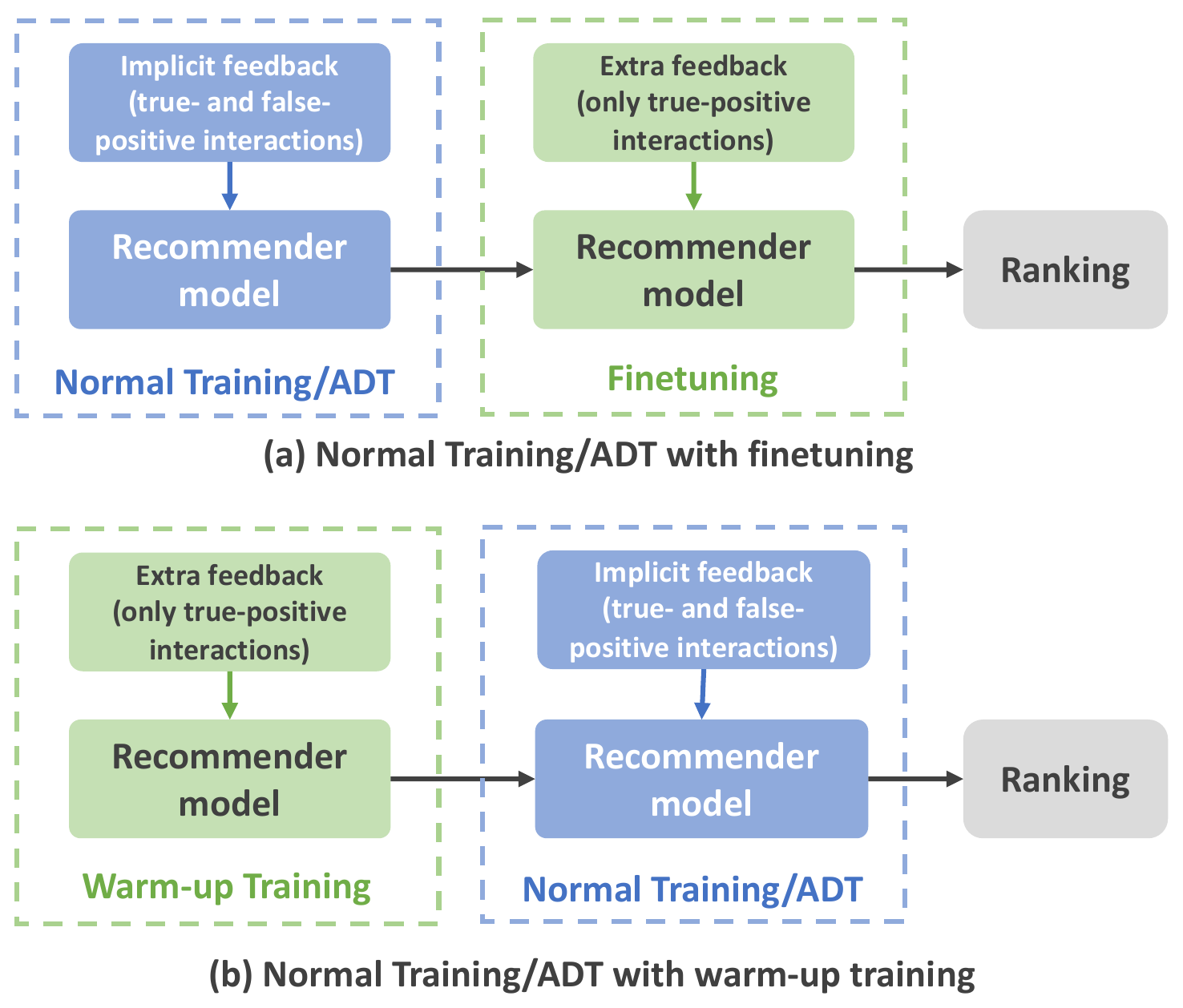}
\caption{Illustration of finetuning and warm-up training with extra feedback.}
\label{fig:finetuning}
\end{figure}

\subsection{ADT with Extra Feedback}

Although extra feedback (\eg ratings) is usually sparse, it is reliable to reflect the actual user satisfaction, \ie indicating the true-positive interactions. We thus further utilize extra feedback for ADT when available\footnote{This is not equal to clean training in Section \ref{sec:study} because available extra feedback is insufficient to identify all true-positive interactions in implicit feedback.}. 

\subsubsection{\textbf{Training with Extra Feedback}}
By considering the order of training with implicit feedback and extra feedback, we introduce two training strategies: finetuning and warm-up training.

\vspace{5pt}
\noindent{\textbf{Finetuning.}}
As shown in Figure \ref{fig:finetuning}(a), an intuitive idea is to retrain the recommender model by fitting the available true-positive interactions. Specifically, we first train the recommender model with ADT over the implicit feedback, and finetune the model over the extra feedback. The finetuning will refine the recommendation lists of users based only on the extra feedback. Formally, given the extra feedback $D^*$, the model parameter $\Theta^*$ obtained from ADT will be further optimized by:
\begin{equation}\label{eqn:finetuning}
    \arg\min_{\Theta^*} \mathcal{L}_{CE}(u, i), \textit{ where } (u, i, {y}^*_{ui})\in D^*,
\end{equation}
where the negative interactions in $D^*$ are sampled with the help of implicit feedback $\bar{D}$. Only the interactions which are not in $\bar{D}$ are sampled into $D^*$ to partly avoid using the missing true-positive interactions as negative ones. 

\vspace{5pt}
\noindent{\textbf{Warm-up Training.}}
From Section \ref{sec:observations}, we know that the recommender model learns to distinguish the true-positive and false-positive interactions in the early stage of training. If extra feedback $D^*$ is available, we can leverage it to ``warm up'' the recommender model before ADT (illustrated in Figure \ref{fig:finetuning}(b)). Warm-up training will prevent the recommender model from being misled by false-positive interactions in the beginning of training.
The recommender model will thus be more reliable to identify false-positive interactions by the large loss values during ADT. Formally, warm-up training can be seen as optimizing the model initialization for ADT by Equation (\ref{eqn:finetuning}).

\subsubsection{\textbf{Colliding Inference}}
The key to the success of incorporating extra feedback lies in resolving its sparsity issue. 
For the users with sparse extra feedback, the influence of extra feedback $D^*$ on their final prediction scores $\mathbf{\hat{y}}$ is weakened due to the small number of extra feedback as opposed to implicit feedback. Therefore, our target is to enhance the effect of reliable $D^*$ on $\mathbf{\hat{y}}$.

\vspace{5pt}
\noindent\textbf{Implementation.} 
We achieve the target during model inference by supplementing the model prediction for sparse user $u$ with similar users. Formally,
\begin{equation}
\label{eqn:colliding_fusion}
\begin{aligned}
\mathbf{\hat{y}^{'}}_u = \lambda * \mathbf{\hat{y}}_u + (1-\lambda) * \sum_{j\in \mathcal{N}_u}  w_j\mathbf{\hat{y}}_{j},
\end{aligned}
\end{equation}
where $\mathbf{\hat{y}}_u = [\hat{y}_{u1}, ..., \hat{y}_{u|\mathcal{I}|}]^T$ denotes the model prediction of user $u$ over all items (\ie item set $\mathcal{I}$), $\mathcal{N}_u$ is the neighbor set of user $u$, $\mathbf{\hat{y}}_j$ represents the prediction scores of neighbor $j$, and $w_j$ denotes the weights to adjust the contribution of each neighbor. 
Specifically, we calculate the user similarity via inner product in the user representations space learned from extra feedback, and then select $|\mathcal{N}_u|$ similar neighbors for each user. The weighted sum of prediction scores of the neighbors summarizes the colliding effect of extra feedback on the final prediction score $\mathbf{\hat{y}}_u$ (see the following section for the explanation). 
Besides, a hyper-parameter $\lambda$ is employed to balance the original prediction scores $\mathbf{\hat{y}}_u$ and the colliding effect. $\mathbf{\hat{y}^{'}}_u$ is the final ranking score for making recommendation. 

Intuitively, the neighbors are obtained from a user representation space learned from true-positive interactions, where the interactions reflect the actual user preference of sparse users. 
Due to the sparsity of extra feedback, some interests of sparse users are not reflected by the representation, making the representation unsuitable for making prediction. Nevertheless, such representation reflects user similarity regarding their true interests.
As such, we lookup neighbors according to the representation and augment the recommendations of the sparse user with neighbors' recommendations.

\begin{figure}[t]
\centering
\setlength{\abovecaptionskip}{0cm}
\includegraphics[scale=0.5]{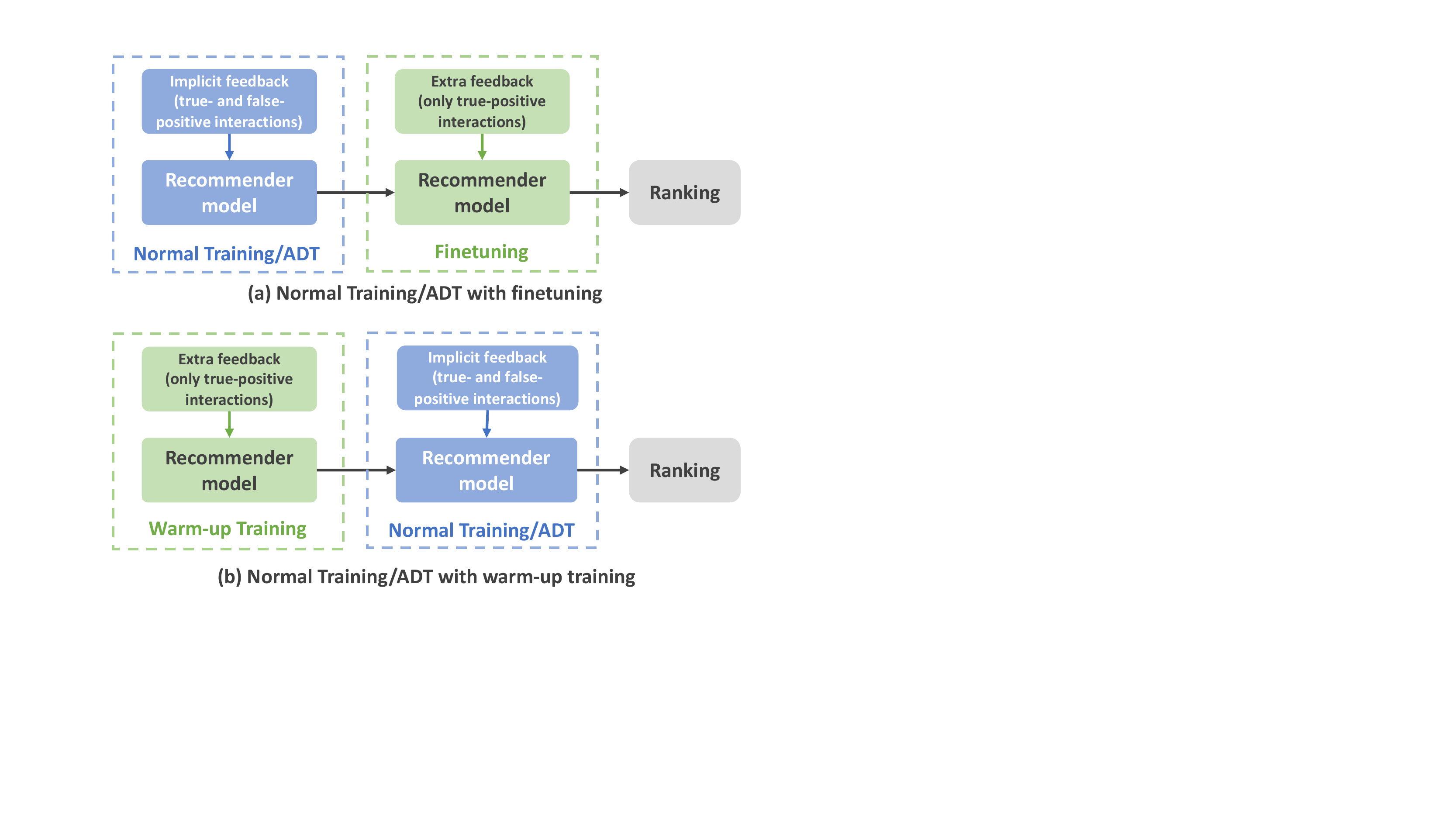}
\caption{Causal graph for ADT with warm-up training.}
\label{fig:colliding}
\vspace{-0.2cm}
\end{figure}

\vspace{5pt}
\noindent\textbf{Rationality.}
The idea of Equation~\ref{eqn:colliding_fusion} comes from the colliding effect in causality theory~\cite{pearl2009causality}. In this work, we utilize warm-up training as an example to illustrate the underlying theory of colliding inference because of its superior performance to enhance ADT (see the empirical evidence in Section \ref{sec:adt_extra}). 


As shown in Figure \ref{fig:colliding}, we introduce a causal graph~\cite{pearl2009causality, Pearl2018the} to inspect the causal relations in ADT with warm-up training for sparse users, where nodes and edges represent the variables and causal relations, respectively~\cite{wang2020click}. Specifically, it presents extra feedback ($D^*$), implicit feedback ($\bar{D}$), user representations after warm-up training ($E^*$), user representations after ADT ($\hat{E}$), prediction scores over items after warm-up training ($Y^*$), and prediction scores over items after ADT ($\hat{Y}$). 
In particular, 
\begin{itemize}
    \item $D^* \rightarrow E^*$ or $\bar{D} \rightarrow \hat{E}$: the user representations are learned from the user feedback. 
    \item $\bar{D} \rightarrow E^*$: the user representations after warm-up training are affected by implicit feedback because it is used to sample negative interactions during training.
    \item $E^* \rightarrow Y^*$ or $\hat{E} \rightarrow \hat{Y}$\footnote{We assume that $E^*$ and $\hat{E}$ are independent for the users with sparse extra feedback. While $E^*$ is the initialization of $\hat{E}$, the influence of the initialization $E^*$ will be easily overwhelmed by the extensive implicit feedback during ADT~\cite{Hu_2021_CVPR}.}: the prediction scores are based on the user representations.
   
\end{itemize}

According to the $d$-separation principle~\cite{pearl2009causality}, $D^*$ is independent from $\bar{D}$, $\hat{E}$, and $\hat{Y}$, which blocks the effect of extra feedback $D^*$ on the final prediction score $\hat{Y}$ of sparse users. Considering that $E^*$ is a collider between $D^*$ and $\bar{D}$,
conditioning on E* will make $D^*$ and $\hat{Y}$ conditionally $d$-connected~\cite{pearl2009causality,Hu_2021_CVPR}. In this way, we can estimate the colliding effect~\cite{Pearl2018the} of $D^*$ on $\hat{Y}$, which is formulated as:,
\begin{equation}
\label{eqn:colliding_effect}
\begin{aligned}
&P(\hat{Y}| {E}^*, D^* = d_u) - P(\hat{Y}| {E}^*, D^* = \varnothing) \\
&= \sum_{\bar{D}} P(\hat{Y}|{E}^*, \bar{D})(P(\bar{D}|{E}^*, D^* = d_u) -P(\bar{D}|{E}^*, D^* = \varnothing)) \\
&= \sum_{\bar{D}} P(\hat{Y}|\bar{D})(P(\bar{D}|{E}^*, D^* = d_u) -P(\bar{D}|{E}^*, D^* = \varnothing)), \\
\end{aligned}
\end{equation}
where $D^* = \varnothing$ denotes the reference status without any extra feedback~\cite{wang2020click, Hu_2021_CVPR} and $d_u$ indicates the true-positive interactions of user $u$. $P(\hat{Y}|{E}^*, \bar{D}) = P(\hat{Y}|\bar{D})$ because ${E}^*$ can only affect $\hat{Y}$ via $\bar{D}$. Intuitively, the causal effect in Equation (\ref{eqn:colliding_effect}) quantifies the influence of extra feedback of user $u$ on the prediction scores by comparing two statuses. 

We utilize the technique of \textit{Distilling Colliding Effect}~\cite{Hu_2021_CVPR} to calculate Equation (\ref{eqn:colliding_effect}). In particular, we will employ a weighting method to estimate the causal effect. Formally, $P(\bar{D}|{E}^*, D^* = d_u) -P(\bar{D}|{E}^*, D^* = \varnothing)$ indicates the the probability of selecting users' implicit feedback $\bar{D}$ given ${E}^*$ and $D^*$, which can be denoted as the weight $W(\bar{D}, {E}^*, D^*)$~\cite{Hu_2021_CVPR}, corresponding to the $w_j$ in Equation (\ref{eqn:colliding_fusion}). Besides, $P(\hat{Y}|\bar{D})$ represents the prediction scores after ADT given the implicit feedback $\bar{D}$ of one user. 
Because of the large sample space of $\bar{D}$, the sum over $\bar{D}$ is approximated by only selecting the implicit feedback of neighbors $\mathcal{N}_u$ with the highest weights $W(\cdot)$. In this way, Equation (\ref{eqn:colliding_effect}) is estimated by the weighted sum of $P(\hat{Y}|\bar{D})$ over the neighbors, \ie $\sum_{j\in \mathcal{N}_u}  w_j\mathbf{\hat{y}}_{j}$.
Correspondingly, we search the neighbors of each sparse user according to similarity in the space of ${E}^*$, and then refine the user's prediction scores $\hat{Y}$ by combining the prediction scores of the neighbors.

\section{Related Work}
\label{sec:related_work}

Denoising implicit feedback is highly related to the negative experience identification, incorporating various feedback, and the robustness of recommendation. 
Besides, we introduce causal recommendation that is related to colliding inference.

\vspace{5pt}
\noindent\textbf{Negative Experience Identification.}
To mitigate the gap between implicit feedback and the actual user preference, many studies have focused on identifying negative experiences in implicit signals \cite{fox2005Evaluating, jiang2020aspect, Gal2019When}. Prior work usually collects the various users' feedback (\textit{e.g.}, dwell time \cite{Kim2014Modeling}, gaze patterns \cite{Zhao2016Gaze}, and skip \cite{fox2005Evaluating}) and the item characteristics \cite{Lu2018Between, Lu2019effects} to predict the user's satisfaction. 
However, they need additional feedback and extensive manual label work, \eg users have to tell if they are satisfied for each interaction. Besides, the quantification of item quality is non-trivial \cite{Lu2018Between}, which largely relies on the manually feature design  and the labeling of domain experts \cite{Lu2018Between}. The unaffordable labor cost hinders the practical usage of these methods, especially in the scenarios with constantly changing items.

\vspace{5pt}
\noindent\textbf{Incorporating Various Feedback.}
To alleviate the impact of false-positive interactions, previous approaches \cite{liu2010understanding, Yang2012Exploiting} also consider incorporating more feedback (\textit{e.g.}, dwell time~\cite{Yi2014Beyond}, skip~\cite{Wen2019Leveraging}, and adding to favorites) into training directly. For instance, Wen \textit{et al.} \cite{Wen2019Leveraging} proposed to train the recommender using three kinds of items: ``click-complete'', ``click-skip'', and ``non-click'' ones. The last two kinds of items are both treated as negative samples but with different weights. 
However, extra feedback might be unavailable in complex scenarios. 
For example, we cannot acquire dwell time and skip patterns after users buy products or watch movies in a cinema. Most users even don't give any informative feedback after clicks. In an orthogonal direction, this work explores denoising implicit feedback without additional data during training, \ie ADT. Furthermore, if some extra feedback (\eg ratings) is available, we also introduce three strategies to utilize it to enhance ADT.

\vspace{5pt}
\noindent\textbf{Robustness of Recommender Systems.}
Gunawardana \textit{et al.} \cite{gunawardana2015evaluating} defined the robustness of recommender systems as ``the stability of the recommendation in the presence of fake information''. Prior work \cite{Lam2004Shilling, Shriver2019Evaluating} has tried to evaluate the robustness of recommender systems under various external attacks, such as shilling attacks \cite{Lam2004Shilling}, fuzzing attacks \cite{Shriver2019Evaluating}, and fake feedback from fraudsters~\cite{zhang2020gcn}. To build more robust recommender systems, some auto-encoder based models \cite{Florian2015Collaborative, wu2016collaborative, Liang2018Variational} introduce the denoising techniques. 
These approaches (\eg CDAE~\cite{wu2016collaborative}) first corrupt the interactions of user by random noises, and then try to reconstruct the original one with auto-encoders. Besides, adversarial learning~\cite{wang2017irgan} might serve as a reasonable solution to defend the diverse attacks or detect fake users~\cite{cao2020adversarial, chris2019adversarial}. 
And robust learning~\cite{han2018co} will also be a promising choice to empower recommender models with the ability of denoising user feedback~\cite{yu2020sampler}.
However, as these methods focus on external attacks or random noises, they ignore the natural false-positive interactions in data. This work highlights the negative impact of natural noisy interactions, and improve the robustness against them.

\vspace{5pt}
\noindent\textbf{Causal Recommendation.}
Recently, causal inference~\cite{pearl2009causality, Pearl2018the} has been widely used to liberate deep learning from blindly fitting the correlation~\cite{niu2020counterfactual, tang2020unbiased, feng2021empowering, feng2021should}. Theoretically, it leverages causal graphs to describe the causal relationships between different variables and conducts three-level reasoning over the causal graph: observation, intervention, and counterfactual~\cite{pearl2009causality}. As to the information retrieval domain, early research~\cite{ai2018unbiased, bonner2018causal, liang2016modeling, wang2019doubly} starts from using causal inference to remove various bias in user feedback~\cite{ROSENBAUM1983the}, such as position bias~\cite{Marco2020Controlling} and popularity bias~\cite{abdollahpouri2019managing}. Recently, more techniques in causal inference have been applied to solve the issues in recommendation~\cite{Konstantina2020Deconfounding, saito2020unbiased, Liu2020a}, for example, causal intervention~\cite{wang2021deconfounded} and counterfactual inference~\cite{wang2020click}. However, existing work on causal recommendation has never studied the colliding effect or the causality-based inference strategy. 
\section{Experiment}
\label{sec:experiment}

\vspace{2pt}
\textbf{Dataset.}
To evaluate the effectiveness of ADT, we conducted experiments on three benchmarks: Adressa, Amazon-book, and Yelp. Extra feedback (\eg ratings and dwell time) is used to identify the false-positive interactions. 
\begin{itemize}[leftmargin=*]
    \item \textbf{Adressa:} This is a news recommendation dataset from Adressavisen\footnote{\url{https://www.adressa.no/}}~\cite{Gulla2017the}. It includes user clicks over news and the dwell time for each click, where the clicks with dwell time < 10s are treated as false-positive ones~\cite{Yi2014Beyond, Kim2014Modeling}.
    \item \textbf{Amazon-book:} It is from the Amazon-review datasets\footnote{\url{http://jmcauley.ucsd.edu/data/amazon/}}~\cite{he2016ups}. It records users' purchases over books with rating scores. A rating score below 3 is regarded as a false-positive interaction.
    \item \textbf{Yelp\footnote{\url{https://www.yelp.com/dataset/challenge}}:} It covers the businesses in the catering industry (\eg restaurants and bars), which are reviewed by users. Similar to Amazon-book, the rating scores below 3 are false-positive interactions.
\end{itemize}

We split the dataset into training, validation, and testing sets, and explored two experimental settings:
\begin{itemize}[leftmargin=*]
    \item[1)] Extra feedback is unavailable during training (Section \ref{sec:p_com} and \ref{sec:indepth}). To evaluate the performance of denoising implicit feedback, we kept all interactions, including the false-positive ones, in training and validation, and tested the models only on true-positive interactions. 
    \item[2)] Sparse extra feedback is available during training (Section \ref{sec:adt_extra}). We assume that partial true-positive interactions have already been known, which will be used to verify the performance of the proposed three strategies: finetuning, warm-up training, and colliding inference. 
\end{itemize}

%
%
\begin{table}[t]\small
\setlength{\abovecaptionskip}{0cm}
\setlength{\belowcaptionskip}{0cm}
\caption{Statistics of the datasets. \#FP interactions refer to the number of false-positive interactions.}
\label{table2}
\centering
\setlength{\tabcolsep}{1.2mm}{
\begin{tabular}{l|l|l|l|l}
\hline
\textbf{Dataset}      & \textbf{\#User} & \textbf{\#Item} & \textbf{\#Interaction} & \textbf{\#FP Interaction}       \\ \hline
Adressa & 212,231  & 6,596   & 419,491        & 247,628            \\ \hline
Amazon-book  & 80,464  & 98,663  & 2,714,021      & 199,475             \\ \hline
Yelp         & 45,548  & 57,396  & 1,672,520      & 260,581             \\ \hline
\end{tabular}
}
\end{table}
%
%

%
%
\begin{table*}[t]
\setlength{\abovecaptionskip}{0cm}
\caption{Overall performance of three testing recommender models trained with ADT strategies and normal training over the three datasets. Note that Recall@K and NDCG@K are shorted as R@K and N@K to save space, respectively, and ``RI'' in the last column denotes the relative improvement of ADT over normal training on average. The best results are highlighted in bold.}
\label{table3}
\centering
\resizebox{\textwidth}{!}{
\begin{tabular}{l|cccc|cccc|cccc|c}
\hline
\textbf{Dataset} & \multicolumn{4}{c|}{\textbf{Adressa}} & \multicolumn{4}{c|}{\textbf{Amazon-book}} & \multicolumn{4}{c|}{\textbf{Yelp}} &  \\ 
\textbf{Metric} & \textbf{R@3} & \textbf{R@20} & \textbf{N@3} & \textbf{N@20} & \textbf{R@50} & \textbf{R@100} & \textbf{N@50} & \textbf{N@100} & \textbf{R@50} & \textbf{R@100} & \textbf{N@50} & \textbf{N@100} & \textbf{RI} \\ \hline \hline
GMF & 0.0880 & 0.2141 & 0.0780 & 0.1237 & 0.0609 & 0.0949 & 0.0256 & 0.0331 & 0.0840 & 0.1339 & 0.0352 & 0.0465 & -\\ 
GMF+T-CE & \textbf{0.0892} & \textbf{0.2170} & \textbf{0.0790} & \textbf{0.1254} & \textbf{0.0707} & \textbf{0.1113} & \textbf{0.0292} & \textbf{0.0382} & \textbf{0.0871} & \textbf{0.1437} & 0.0359 & \textbf{0.0486} & 7.14\% \\ 
GMF+R-CE  & 0.0891 & 0.2142 & 0.0765 & 0.1229 & 0.0682 & 0.1075 & 0.0275 & 0.0362 & 0.0861 & 0.1361 & \textbf{0.0366} & 0.0480 & 4.34\% \\ \hline
NeuMF & 0.1416 & 0.3159 & 0.1267 & 0.1886 & 0.0512 & 0.0829 & 0.0211 & 0.0282 & 0.0750 & 0.1226 & 0.0304 & 0.0411 & -\\ 
NeuMF+T-CE & \textbf{0.1418} & 0.3106 & 0.1227 & 0.1840 &  \textbf{0.0725} &  \textbf{0.1158} &  \textbf{0.0289} &  \textbf{0.0385} & \textbf{0.0825} & \textbf{0.1396} & \textbf{0.0323} & \textbf{0.0451} & 15.62\% \\ 
NeuMF+R-CE & {0.1414} & \textbf{0.3185} & {0.1266} & \textbf{0.1896} & {0.0628} & {0.1018} & {0.0248} & {0.0334} & 0.0788 & 0.1304 & 0.0320 & 0.0436 & 8.77\% \\ \hline
CDAE & 0.1394 & 0.3208 & 0.1168 & 0.1808 & 0.0989 & 0.1507 & 0.0414 & 0.0527 & 0.1112 & 0.1732 & 0.0471 & 0.0611 &  - \\ 
CDAE+T-CE & \textbf{0.1406} & \textbf{0.3220} & 0.1176 & \textbf{0.1839} & \textbf{0.1088} & \textbf{0.1645} & \textbf{0.0454} & \textbf{0.0575} & \textbf{0.1165} & \textbf{0.1806} & \textbf{0.0504} & \textbf{0.0652} & 5.36\% \\ 
CDAE+R-CE & 0.1388 & 0.3164 & \textbf{0.1200} & 0.1827 & 0.1022 & 0.1560 & 0.0424 & 0.0542 & 0.1161 & 0.1801 & 0.0488 & 0.0632 & 2.46\% \\ \hline
\end{tabular}
}
\end{table*}
%
%

\vspace{5pt}
\noindent\textbf{Evaluation Protocols.}
For each user in the testing set, we predicted the preference score over all the items except the positive ones used during training. Following existing studies \cite{He2017Neural, wang2019NGCF}, we reported the performance \wrt two widely used metrics: Recall@K and NDCG@K, where higher scores indicate better performance. For both metrics, we set K as 50 and 100 for Amazon-book and Yelp, while 3 and 20 for Adressa due to its much smaller item space.

\vspace{5pt}
\noindent\textbf{Testing recommender models.}
To demonstrate the effectiveness of our proposed ADT strategy on denoising implicit feedback, we compared the performance of recommender models trained with T-CE or R-CE and normal training with standard CE. We selected two representative user-based neural models, GMF and NeuMF~\cite{He2017Neural}, and one item-based model, CDAE~\cite{wu2016collaborative}, which is a representative recommender model which can defend random noises in implicit feedback. More details can be found in \cite{wang2021denoising}.


\vspace{5pt}
\noindent\textbf{Parameter Settings.}
For the three testing recommender models, we followed their default settings, and verified the effectiveness of our methods under the same conditions. For GMF and NeuMF, the factor numbers of users and items are both 32. As to CDAE, the hidden size of MLP is set as 200. In addition, Adam~\cite{kingma2014adam} is applied to optimize all the parameters with the learning rate initialized as 0.001 and he batch size set as 1,024. As to the ADT strategies, they have three hyper-parameters in total: $\alpha$ and $\epsilon_{max}$ in the T-CE loss, and $\beta$ in the R-CE loss. In detail, $\epsilon_{max}$ is searched in $\{0.05, 0.1, 0.2\}$ and $\beta$ is tuned in $\{0.05, 0.1, ..., 0.25, 0.5, 1.0\}$. As for $\alpha$, we controlled its range by adjusting the iteration number $\epsilon_{N}$ to the maximum drop rate $\epsilon_{max}$, and $\epsilon_{N}$ is adjusted in $\{1k, 5k, 10k, 20k, 30k\}$. In colliding inference, the number of neighbors $\mathcal{N}_u$ is tuned in $\{1, 3, 5, 10, 20, 50, 100\}$, $w_j$ is set as $1/|\mathcal{N}_u|$, and $\lambda$ is searched in $\{0, 0.1, 0.2, ..., 1\}$. We used the validation set to tune the hyper-parameters and reported the performance on the testing set.

\subsection{Performance Comparison}\label{sec:p_com}
\vspace{2pt}
\noindent\textbf{Overall Performance.}
Table \ref{table3} summarizes the performance of the three testing models trained with standard CE, T-CE, or R-CE. From Table \ref{table3}, we could observe:
\begin{itemize}[leftmargin=*]
    \item In all cases, both the T-CE loss and R-CE loss effectively improve the performance, \eg NeuMF+T-CE outperforms NeuMF by 15.62\% on average over the three datasets. The significant performance gain indicates the better generalization ability of neural recommender models trained by the T-CE loss and R-CE loss. It validates the effectiveness of ADT, \ie discarding or down-weighting hard interactions during training. 
    \item By comparing T-CE and R-CE, we found that the T-CE loss performs better in most cases. We postulate that the recommender model trained with the R-CE loss still suffers from the false-positive interactions, even though they have smaller weights and contribute little to the overall training loss. In addition, the superior performance of the T-CE Loss might be attributed to the additional hyper-parameters in the dynamic threshold function which can be tuned more granularly. Further improvements might be achieved by a fine-grained user-specific or item-specific tuning of these hyper-parameters, which can be done automatically~\cite{lambdaopt}.
    \item Both T-CE and R-CE achieve the largest performance increase on NeuMF, which validates the effectiveness of ADT to prevent vulnerable models from the disturbance of noisy data. On the contrary, the improvement over CDAE is relatively small, showing that the design of defending random noise can also improve the robustness against false-positive interactions to some extent. Nevertheless, applying T-CE or R-CE still leads to performance gain over CDAE, which further validates the rationality of denoising implicit feedback.
\end{itemize}

\noindent In the following, GMF is taken as an example to conduct thorough investigation to save computation cost.



\begin{figure}
\setlength{\abovecaptionskip}{-0.3cm}
  \centering 
  \hspace{-0.25in}
  \subfigure{
    \includegraphics[width=1.8in]{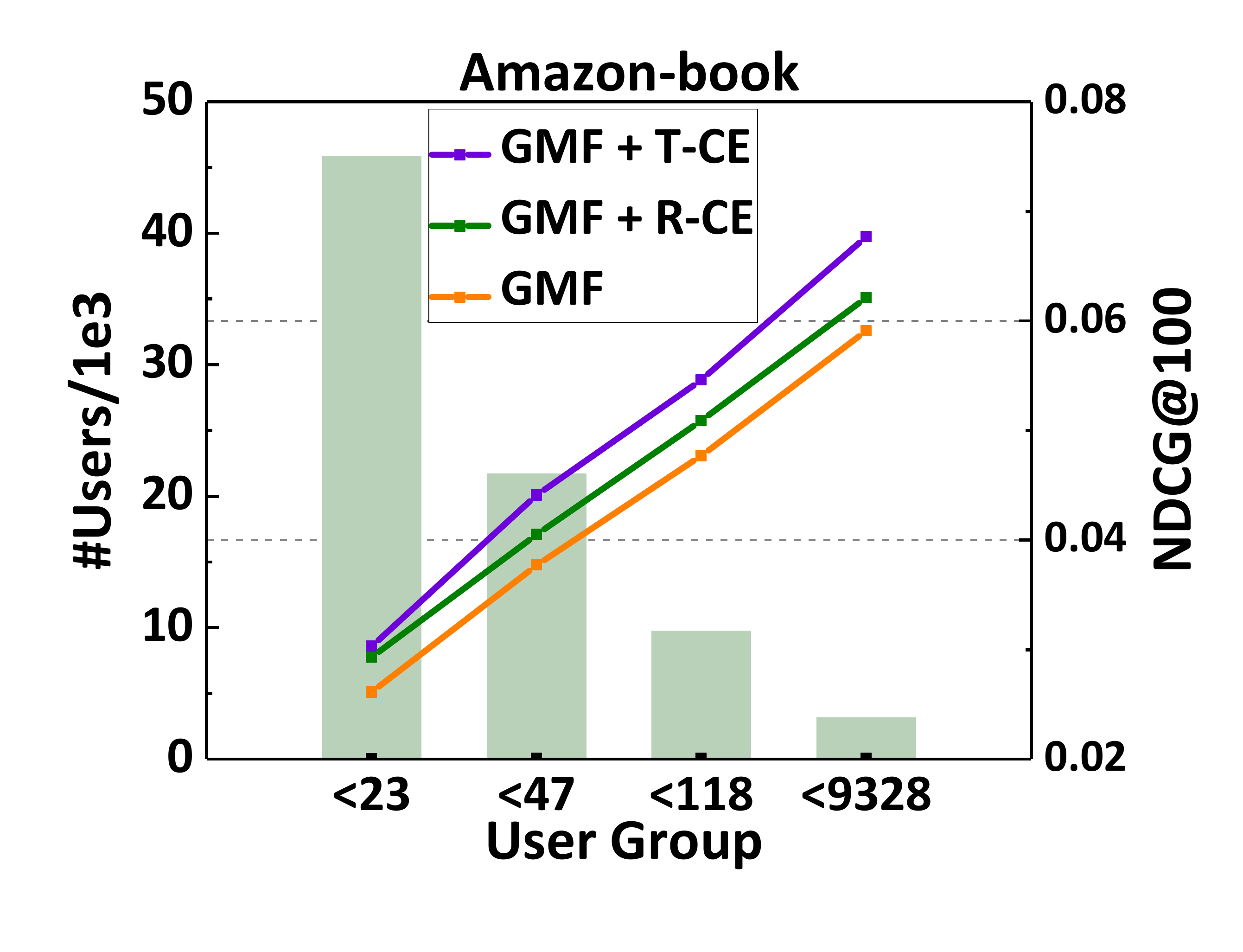}} 
  \hspace{-0.2in}
  \subfigure{
    \includegraphics[width=1.8in]{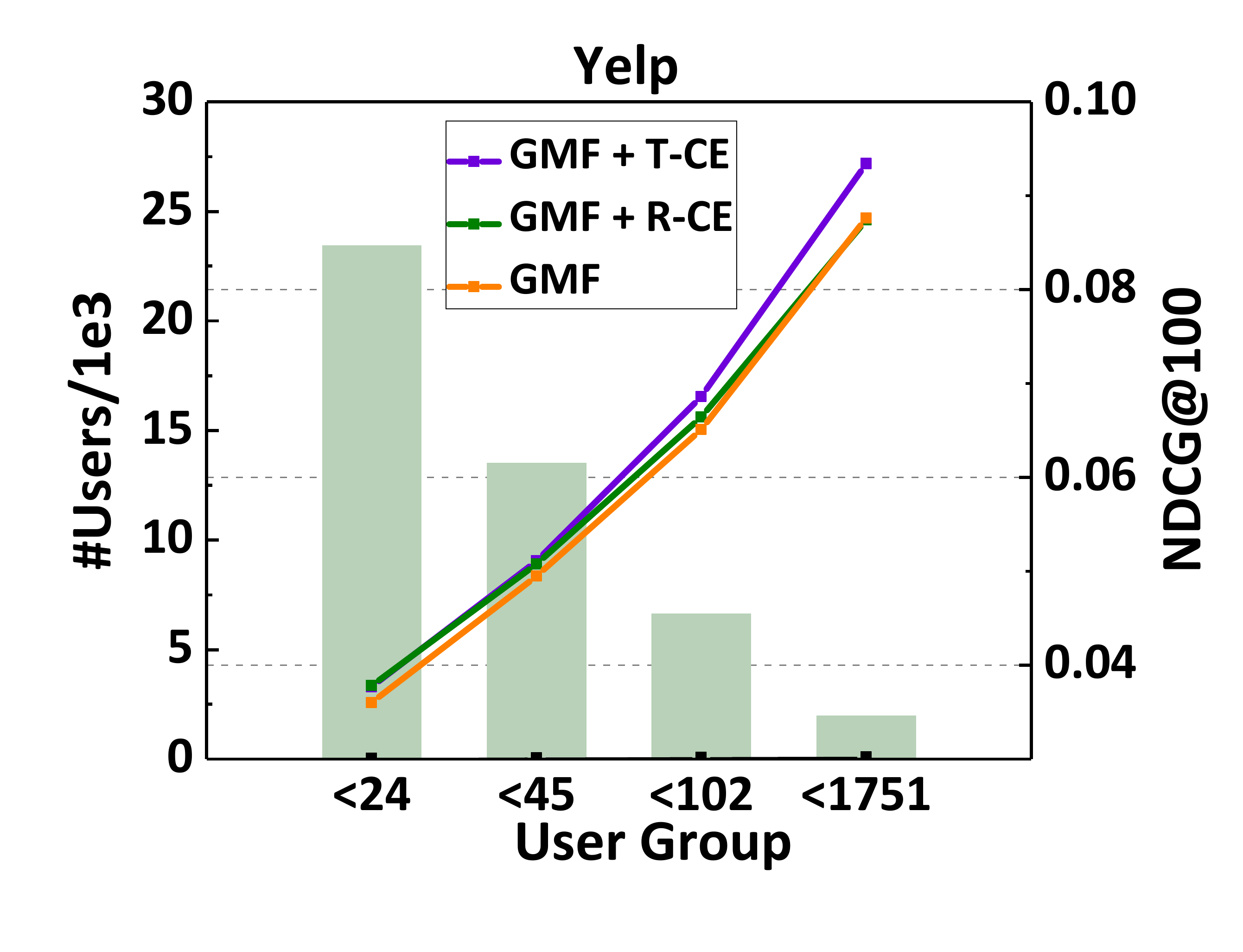}} 
  \hspace{-0.25in} 
  \caption{Performance comparison of GMF over user groups with different sparsity levels. The histograms represent the user number and the lines denote the performance.} 
  \label{fig:usergroup}
  \vspace{-0.2cm}
\end{figure}

\vspace{5pt}
\noindent\textbf{Performance Comparison w.r.t. Interaction Sparsity.}
Since ADT prunes many interactions during training, we explored whether ADT hurts the preference learning of inactive users because their interacted items are sparse. Following the former studies~\cite{wang2019NGCF}, we split testing users into four groups according to the interaction number of each user where each group has the same number of interactions. Figure \ref{fig:usergroup} shows the group-wise performance comparison where we could observe that the proposed ADT strategies achieve stable performance gain over normal training in all cases. It validates that ADT is also effective for the inactive users.

\subsection{In-depth Analysis on ADT}\label{sec:indepth}

\begin{figure}
\setlength{\abovecaptionskip}{0.1cm}
\setlength{\belowcaptionskip}{-0.4cm}
  \centering 
  \hspace{-0.6in}
  \subfigure[Normal Training]{
    \includegraphics[width=1.4in]{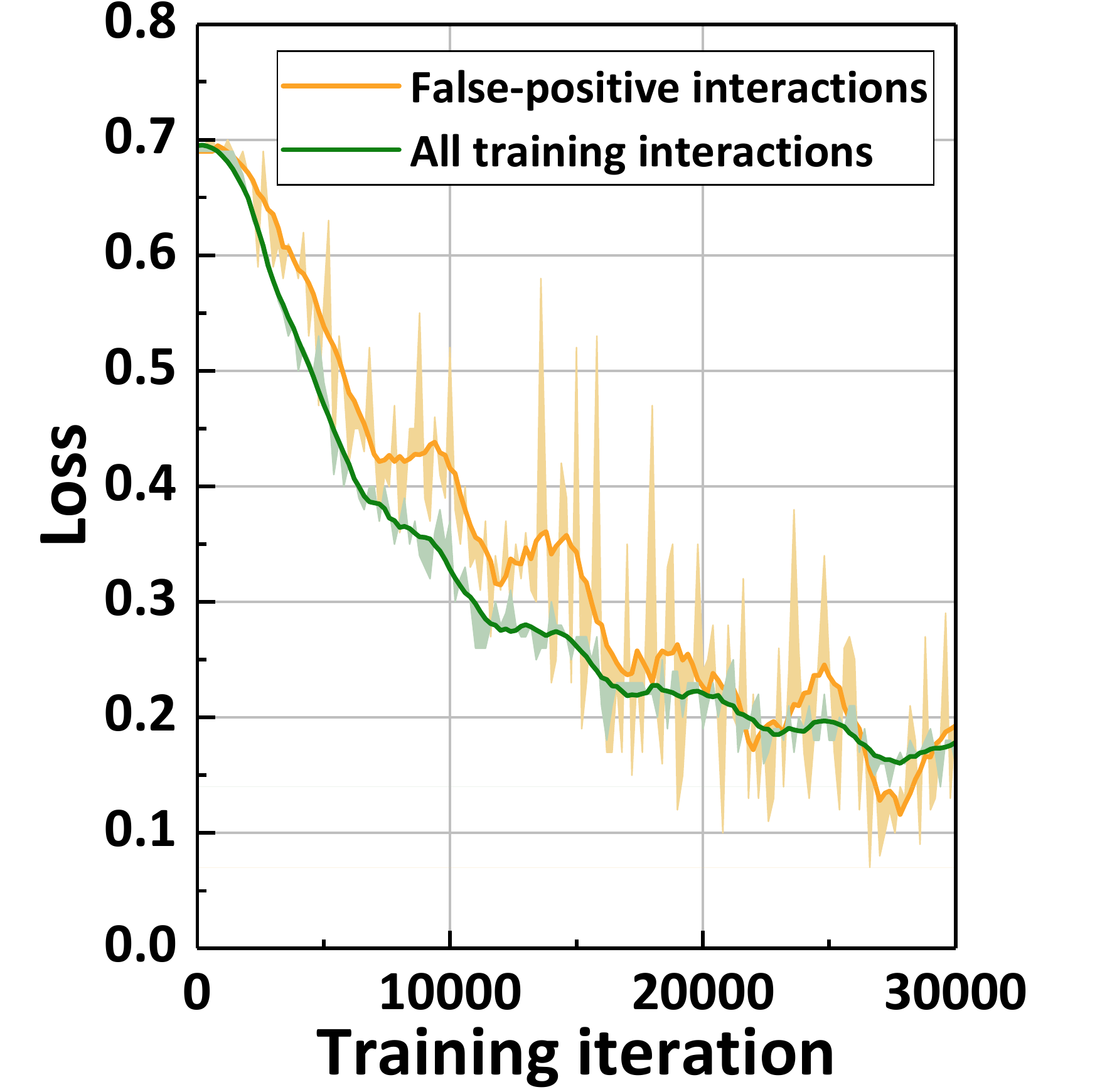}} 
  \hspace{-0.28in}
  \subfigure[Truncated Loss]{
    \includegraphics[width=1.4in]{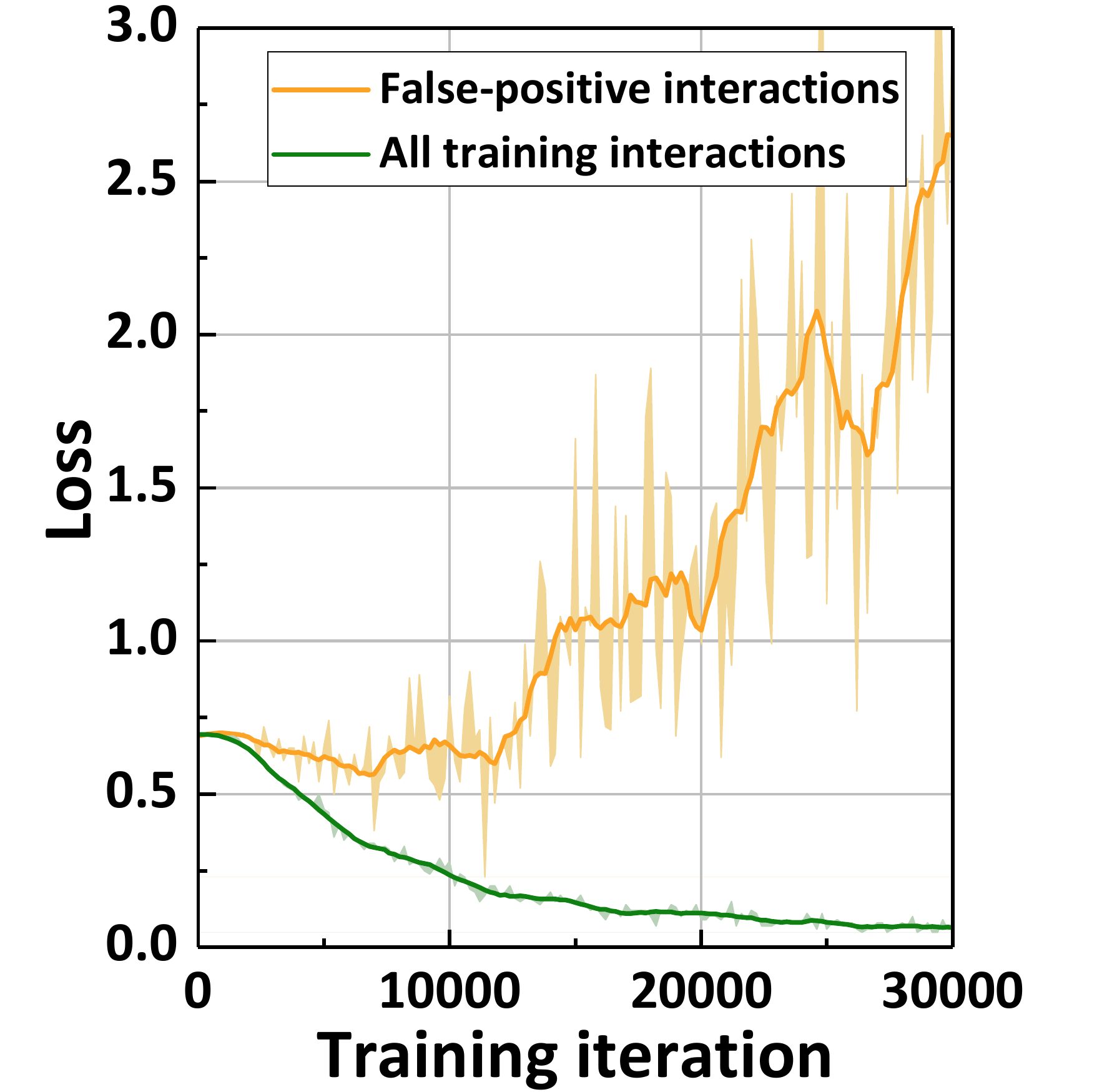}} 
  \hspace{-0.28in}
  \subfigure[Reweighted Loss]{
    \includegraphics[width=1.4in]{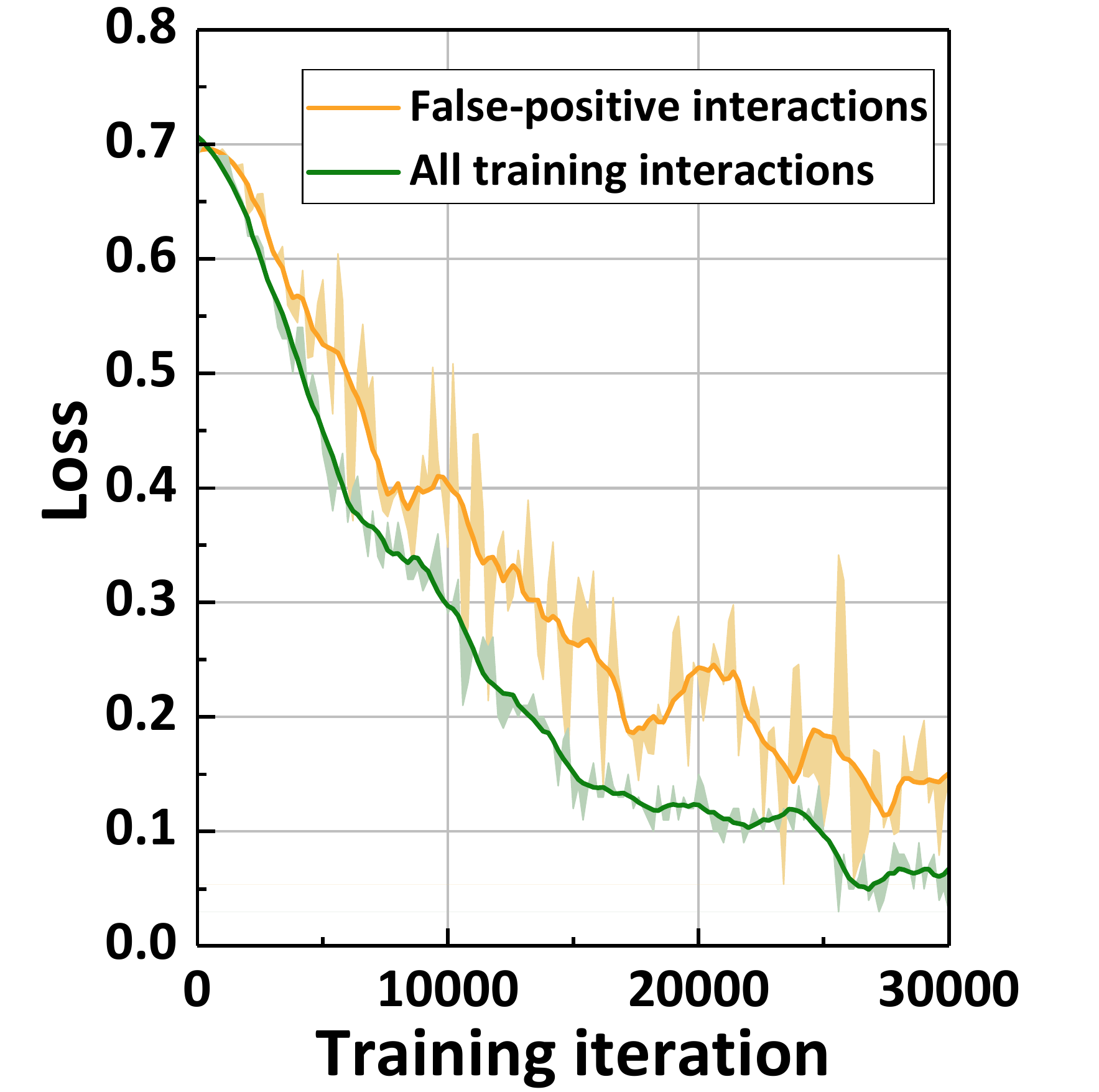}} 
  \hspace{-0.6in} 
  \caption{Loss of GMF with normal training (a), T-CE (b) and R-CE (c).}
  \label{fig:lossTrend}
\end{figure}

\vspace{2pt}
\noindent\textbf{Memorization of False-positive Interactions.}
Recall that false-positive interactions are memorized by recommender models eventually under normal training. We then investigated whether they are also fitted well when ADT is applied. 
From Figure \ref{fig:lossTrend}(a), we could find the CE loss values of false-positive interactions eventually become similar to other interactions, indicating that GMF fits false-positive interactions well at last. On the contrary, as shown in Figure \ref{fig:lossTrend}(b), by applying T-CE, the loss values of false-positive interactions increase while the overall training loss stably decreases step by step. 
This indicates that the model parameters are not optimized over these false-positive interactions, validating the capability of T-CE to identify and discard such interactions. As to R-CE (Figure \ref{fig:lossTrend}(c)), the loss of false-positive interactions also shows a decreasing trend, showing that the recommender model still fits such interactions. However, their loss values are still larger than the real training loss, indicating assigning false-positive interactions with smaller weights is effective. It prevents the model from fitting them quickly. Therefore, we could conclude that both paradigms reduce the effect of false-positive interactions on recommender training, which can explain their improvement over normal training.

\begin{figure}[t]
\vspace{-0.2cm}
\setlength{\abovecaptionskip}{0cm}
\setlength{\belowcaptionskip}{-0cm}
  \centering 
  \hspace{-0.5in}
  \subfigure{
    \includegraphics[width=2in]{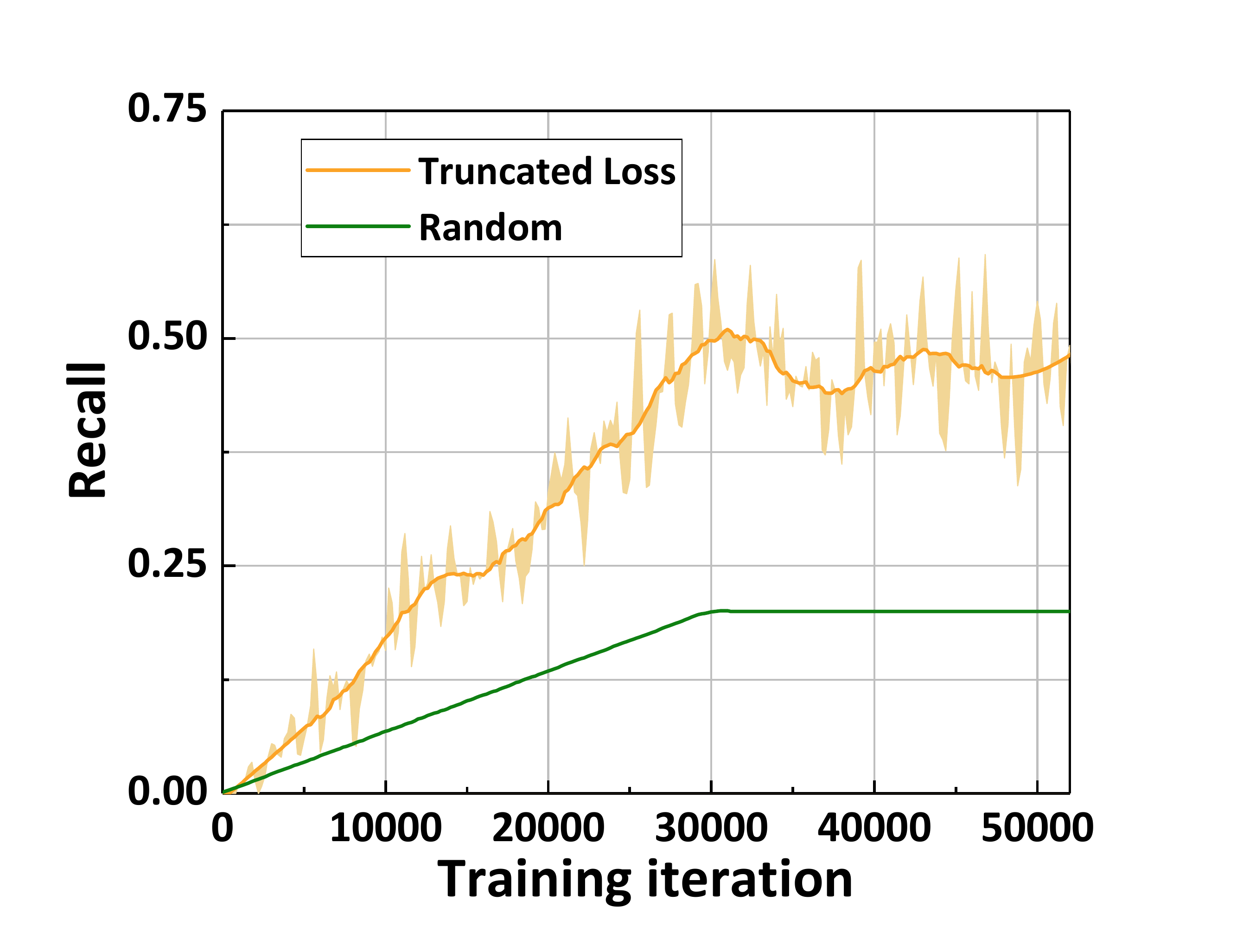}} 
  \hspace{-0.35in}
  \subfigure{
    \includegraphics[width=2in]{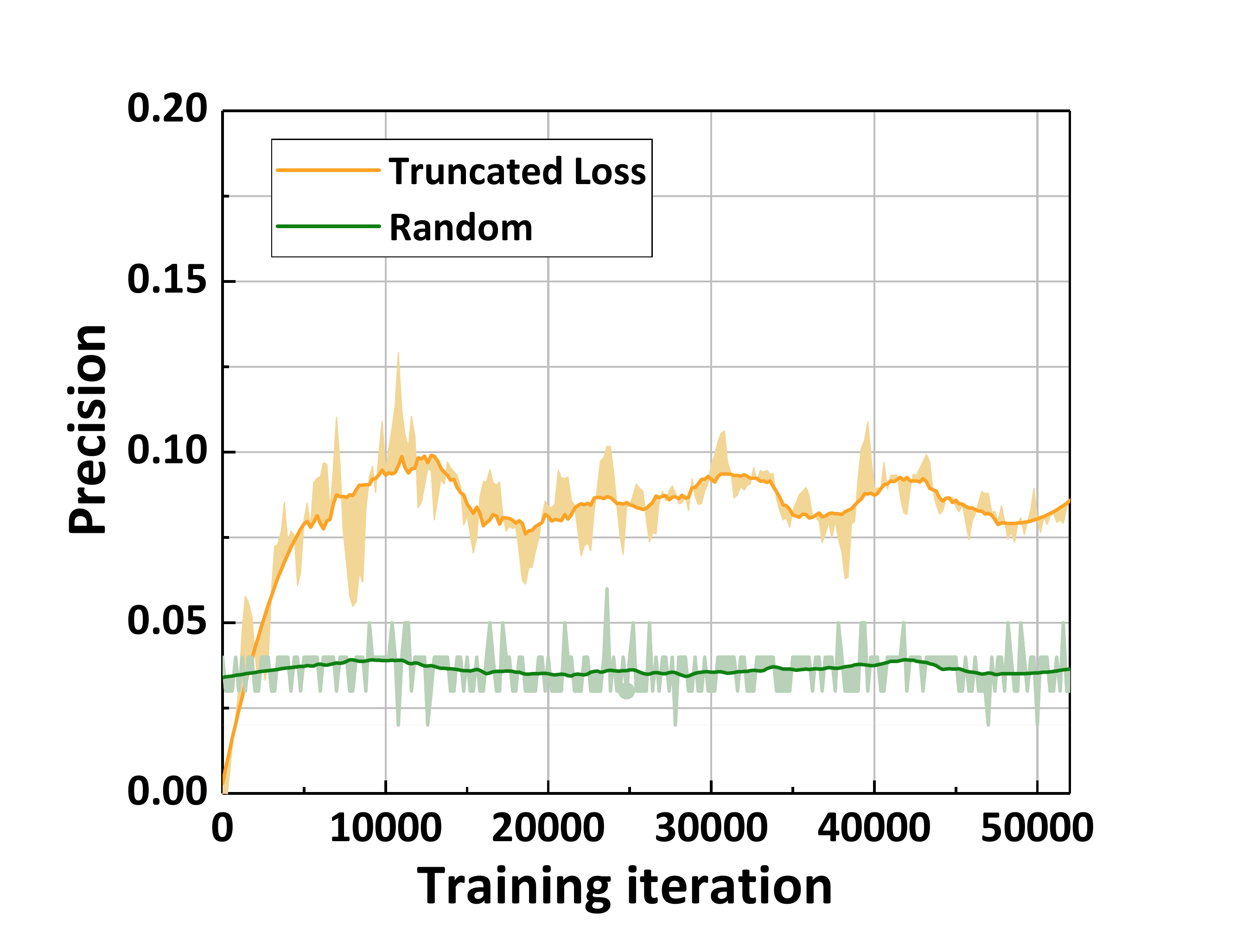}}
  \hspace{-0.5in} 
  \caption{Recall and precision of false-positive interactions over GMF trained with T-CE on Amazon-book.} 
  \label{fig:recall}
\end{figure}

\vspace{5pt}
\noindent\textbf{Study of Truncated Loss.}
Since the Truncated Loss achieves promising performance, we studied how accurately it identifies and discards false-positive interactions. We first defined \textit{recall} to represent what percentage of false-positive interactions in the training data are discarded, and \textit{precision} as the ratio of discarded false-positive interactions to all discarded interactions. Figure \ref{fig:recall} visualizes the changes of the recall and precision along the training process. We take random discarding as a reference, where the recall of random discarding equals to the drop rate $\epsilon(T)$ during training and the precision is the proportion of false-positive interactions within the training batch at each iteration. 
From Figure \ref{fig:recall}, we observed that: 
\begin{itemize}[leftmargin=*]
    \item The Truncated Loss discards nearly half of false-positive interactions after the drop rate keeps stable, greatly reducing the impact of noisy interactions. This might explain the superior performance of T-CE in Table \ref{table3}.
    \item A key limitation of the Truncated Loss is the low precision, \eg only 10\% precision in Figure \ref{fig:recall}, which implies that it inevitably discards many true-positive interactions. However, given the overall superior performance of the Truncated Loss, it seems worthwhile to prune noise interactions even at the cost of losing many true-positive interactions. Nevertheless, how to further improve the precision is a promising research direction in the future. 
\end{itemize}

\begin{figure*}[t]
\setlength{\abovecaptionskip}{0cm}
\setlength{\belowcaptionskip}{0cm}
  \centering 
  \hspace{-0.6in}
  \subfigure{
    \includegraphics[width=1.3in]{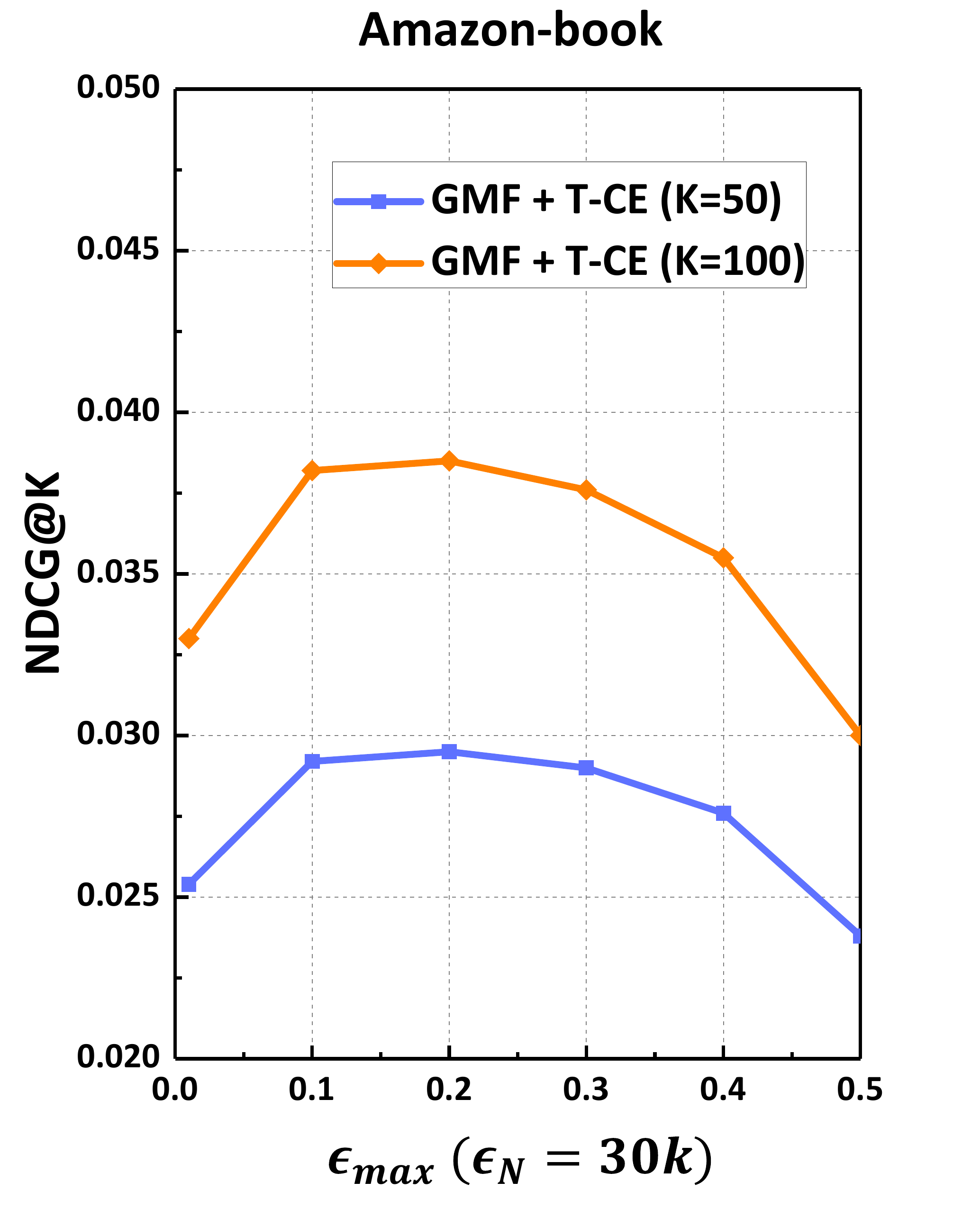}} 
  \hspace{-0.25in}
  \subfigure{
    \includegraphics[width=1.3in]{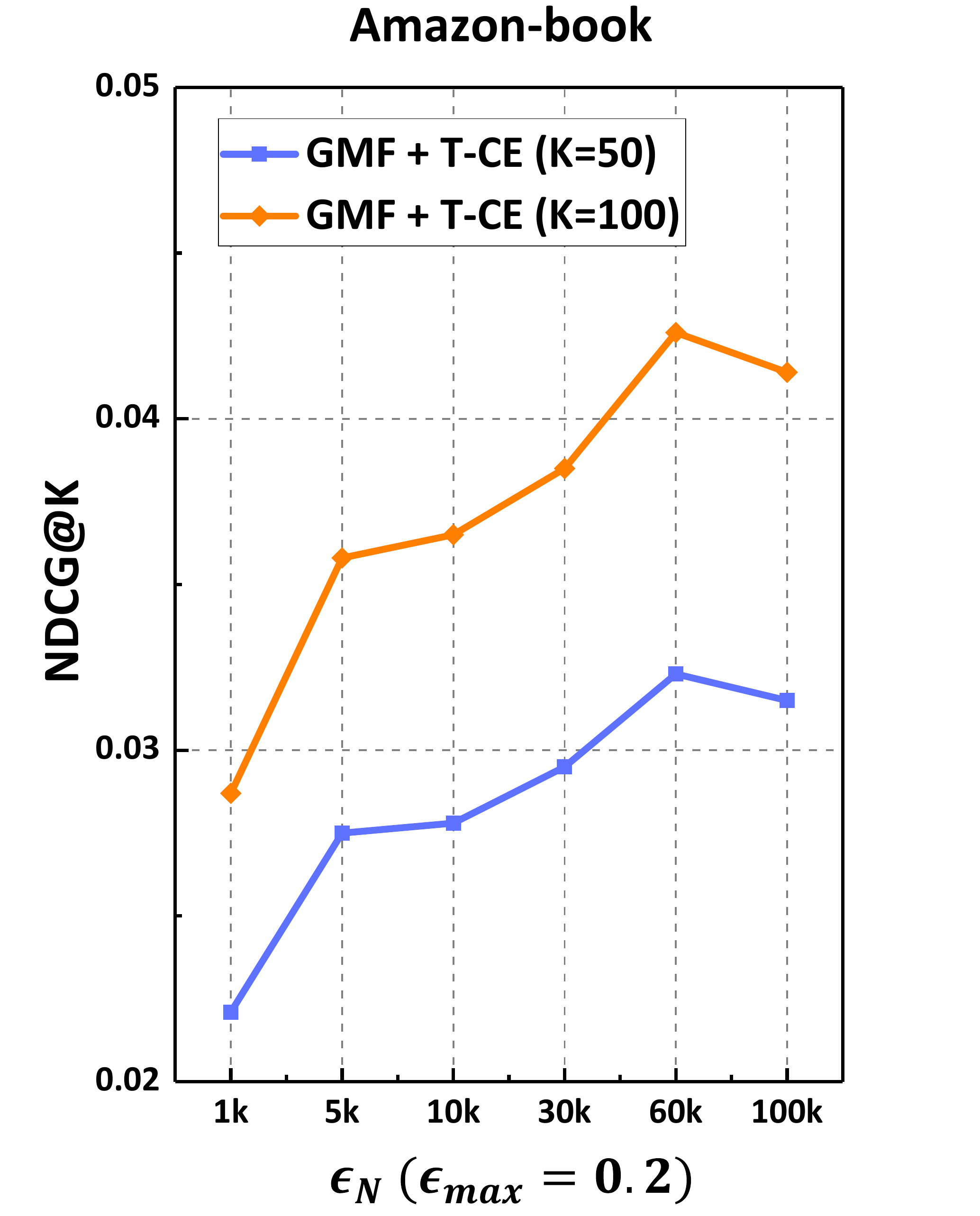}} 
  \hspace{-0.25in}
  \subfigure{
    \includegraphics[width=1.3in]{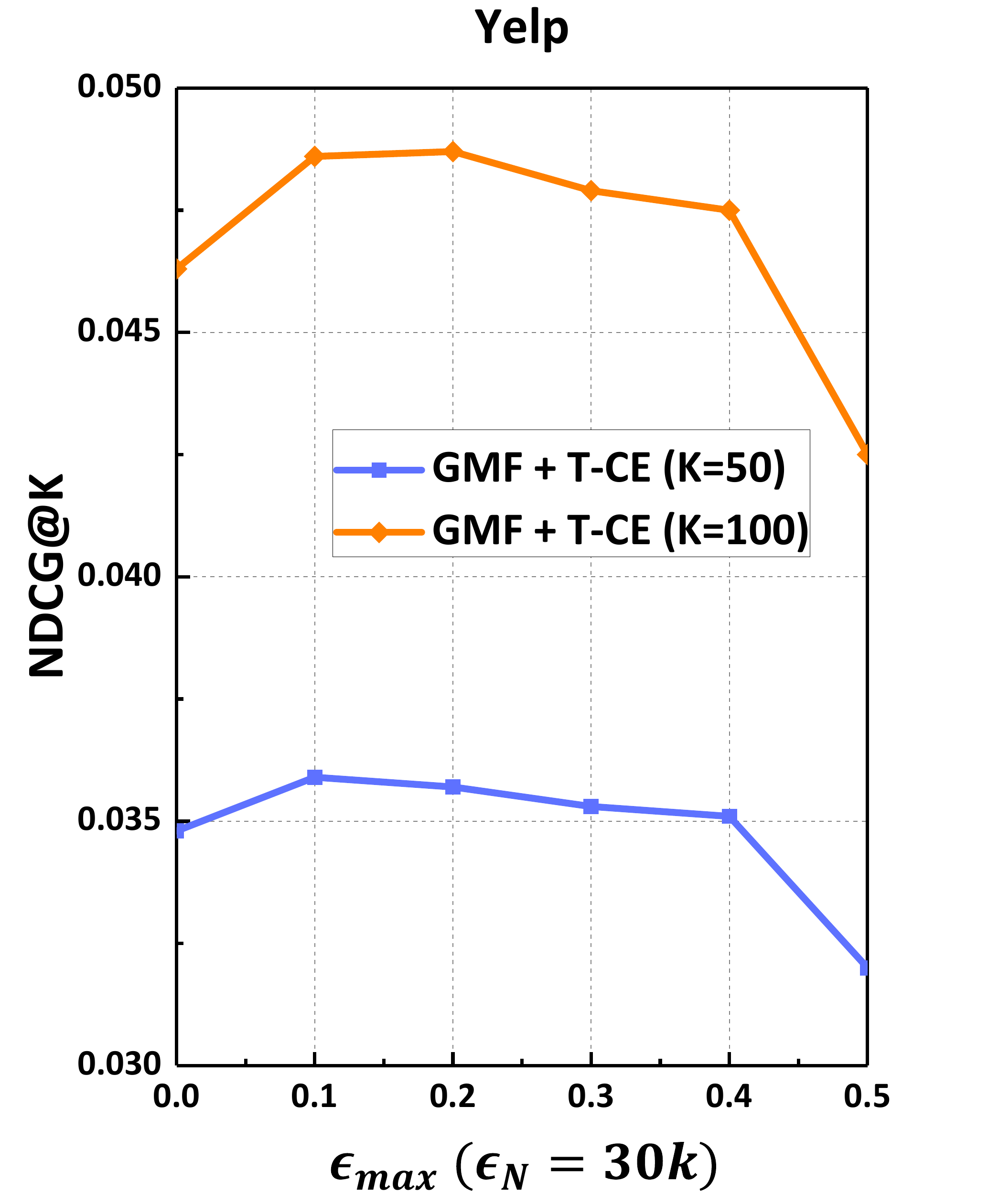}} 
  \hspace{-0.25in}
  \subfigure{
    \includegraphics[width=1.3in]{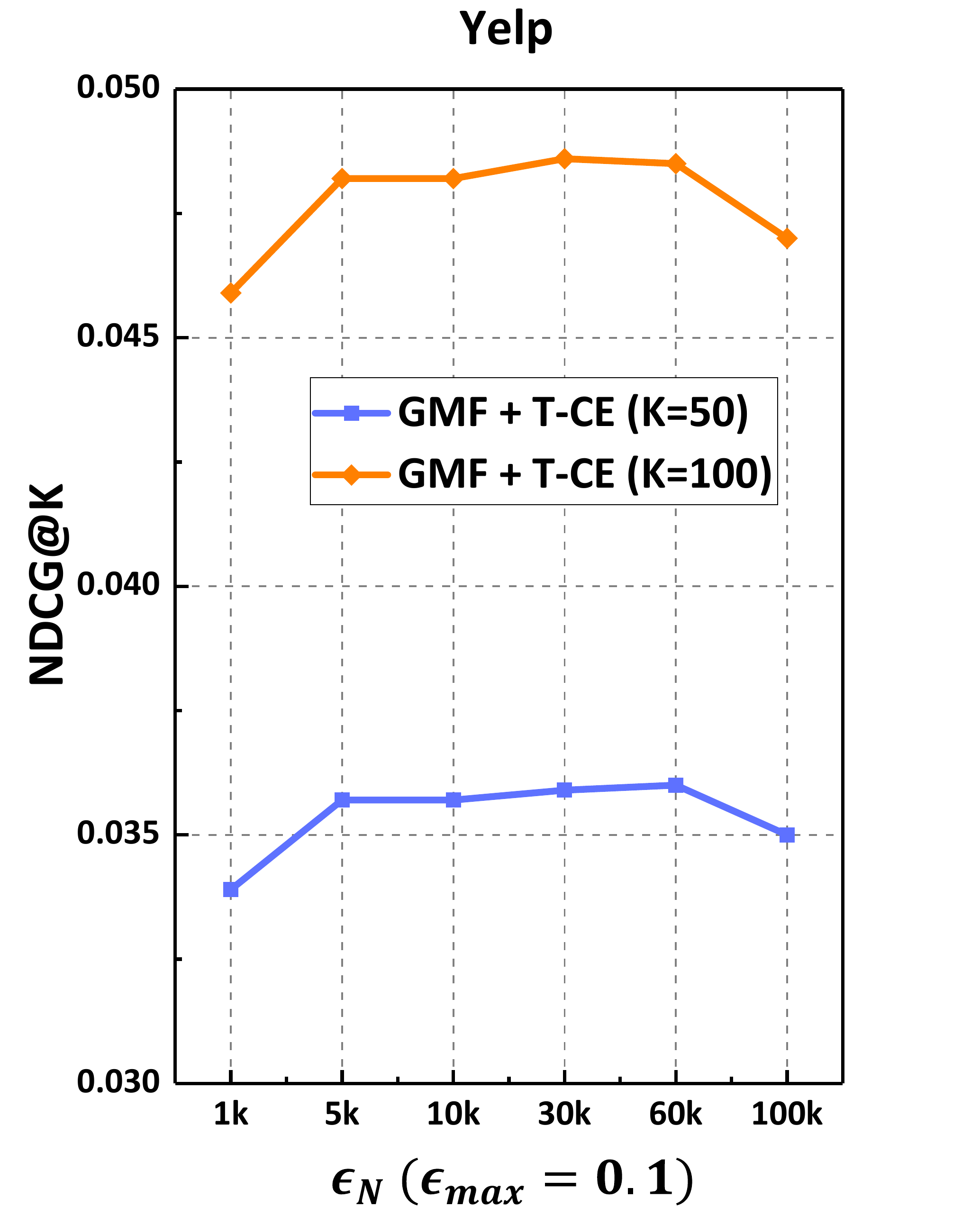}} 
  \hspace{-0.25in}
  \subfigure{
    \includegraphics[width=1.3in]{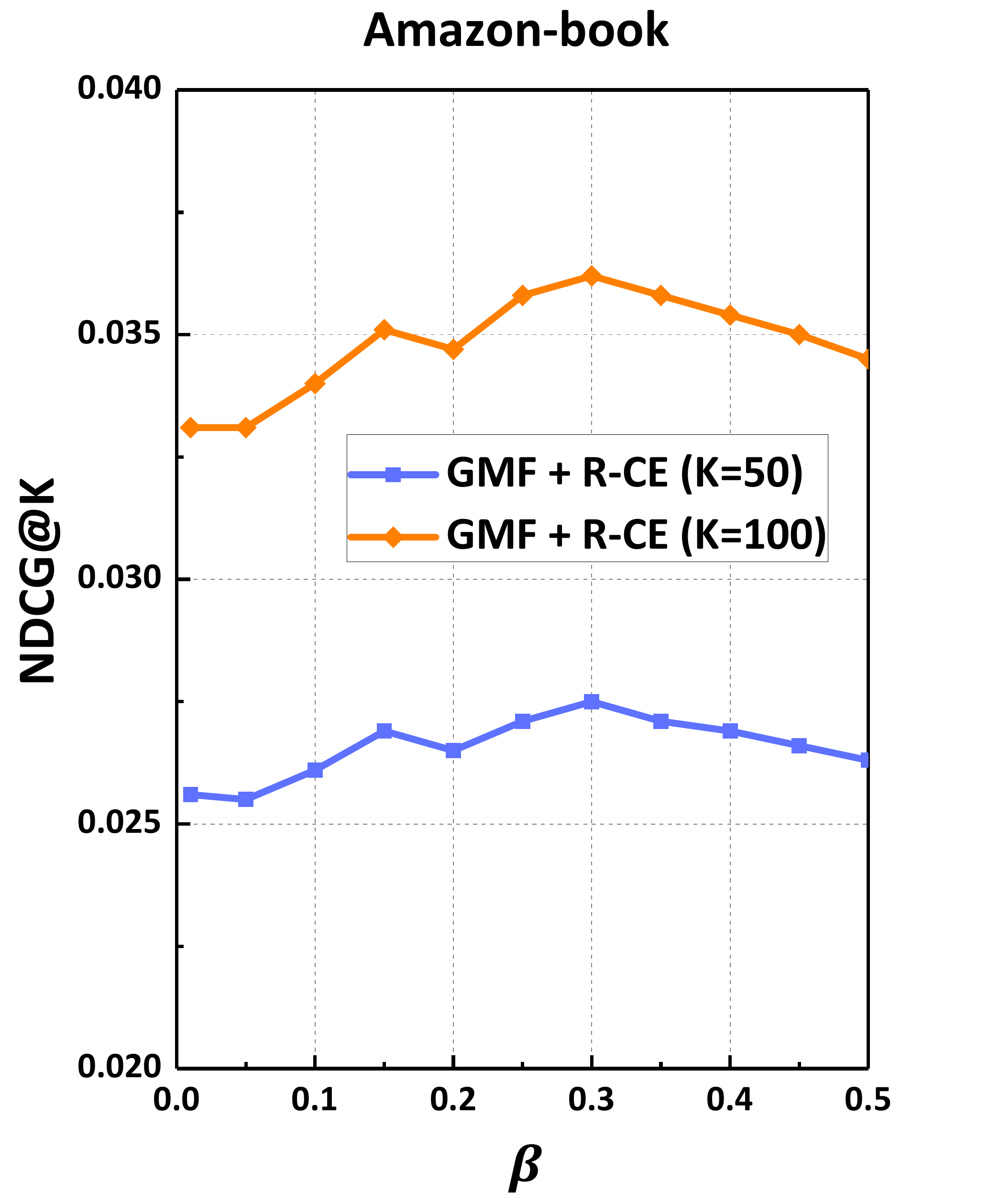}} 
  \hspace{-0.25in}
  \subfigure{
    \includegraphics[width=1.3in]{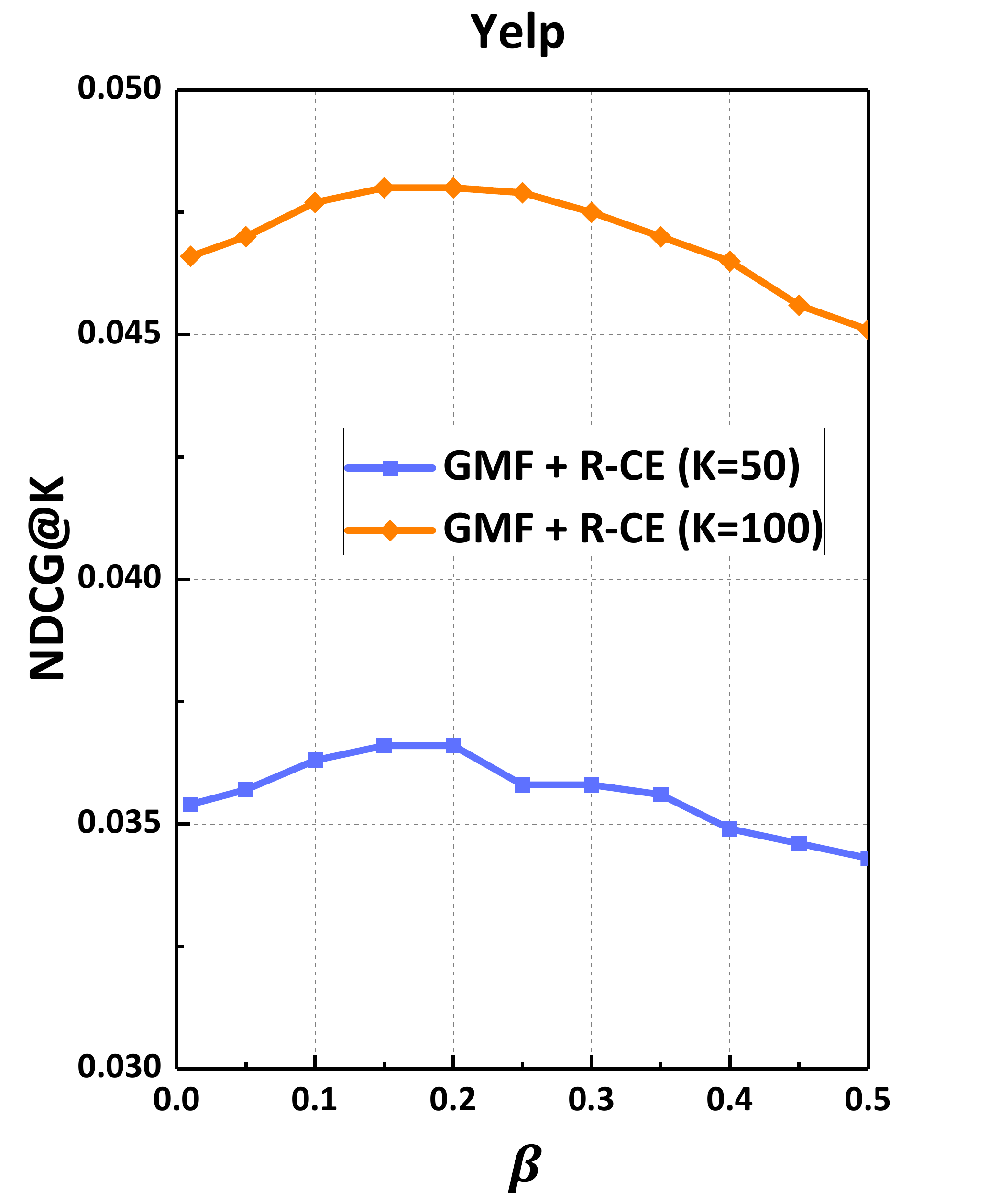}} 
  \hspace{-0.6in} 
  \caption{Performance comparison of GMF trained with ADT on Yelp and Amazon-book \wrt different hyper-parameters, including $\epsilon_{\textit{max}}$, $\epsilon_{N}$, and $\beta$.} 
  \label{fig:parameter}
\end{figure*}

\vspace{5pt}
\noindent\textbf{Hyper-parameter Sensitivity.}
ADT introduces three hyper-parameters into the two paradigms. Specifically, $\epsilon_{max}$ and $\epsilon_{N}$ are used to control the drop rate in T-CE loss, and $\beta$ adjusts the weight function in R-CE loss. To facilitate the future tuning of the two paradigms, we studied how the hyper-parameters affect the performance.
From Figure \ref{fig:parameter}, we could have the following findings: 
\begin{itemize}[leftmargin=*]
    \item The recommender model trained with the T-CE loss performs well when $\epsilon_{max} \in [0.1, 0.3]$. If $\epsilon_{max}$ exceeds 0.4, the performance drops significantly because a large proportion of interactions are discarded. 
    Therefore, the upper bound $\epsilon_{max}$ should be restricted.
    \item The recommender model is relatively sensitive to $\epsilon_{N}$, especially on Amazon-book, and the performance still increases when $\epsilon_{N}$ > 30k. This shows that a limitation of the T-CE loss is the big search space of hyper-parameters. Therefore, it is promising to introduce some automatic hyper-parameter tuning methods~\cite{lambdaopt}, which could further improve denoising performance of the Truncated Loss.
    \item The adjustment of $\beta$ in the Reweighted Loss is consistent across different datasets, and the best results happen when $\beta$ ranges from 0.15 to 0.3. These observations provide insights on how to tune $\beta$ if it's applied to other recommender models and datasets.
\end{itemize}

\subsection{ADT with Extra Feedback}\label{sec:adt_extra}

\begin{table}[]
\setlength{\abovecaptionskip}{0cm}
\caption{Performance of using extra feedback \wrt GMF on Amazon-book. The best results in each method group are highlighted in bold.}
\label{tab:adt_extra}
\centering
\setlength{\tabcolsep}{1.8mm}{
\begin{tabular}{l|cccc}
\hline
\textbf{Method} & \textbf{R@50} & \textbf{R@100} & \textbf{N@50} & \textbf{N@100} \\ \hline \hline
GMF & 0.0609 & 0.0949 & 0.0256 & 0.0331 \\ 
GMF+T-CE & \textbf{0.0707} & \textbf{0.1113} & \textbf{0.0292} & \textbf{0.0382} \\ 
GMF+R-CE & 0.0682 & 0.1075 & 0.0275 & 0.0362 \\ \hline
GMF+NMTR & \textbf{0.0616} & \textbf{0.0967} & \textbf{0.0254} & \textbf{0.0332} \\ 
GMF+NR & 0.0615 & 0.0958 & 0.0254 & 0.0331 \\ \hline
GMF+Finetuning & 0.0608 & 0.0956 & 0.0253 & 0.0330 \\ 
GMF+T-CE+Finetuning & \textbf{0.0722} & \textbf{0.1140} & \textbf{0.0293} & \textbf{0.0386} \\ 
GMF+R-CE+Finetuning & 0.0684 & 0.1087 & 0.0276 & 0.0365 \\ \hline
GMF+Warm-up & 0.0611 & 0.0954 & 0.0255 & 0.0332 \\ 
GMF+T-CE+Warm-up & \textbf{0.0737} & \textbf{0.1159} & \textbf{0.0302} & \textbf{0.0395} \\ 
GMF+R-CE+Warm-up & 0.0691 & 0.1099 & 0.0278 & 0.0368 \\ \hline
\end{tabular}
}
\end{table}

\begin{table*}[]
\setlength{\abovecaptionskip}{0cm}
\caption{Performance of colliding inference over different user groups. The best results in each user group are highlighted in bold.} 
\label{tab:colliding_inference}
\centering
\resizebox{0.8\textwidth}{!}{
\begin{tabular}{l|c|l|cccc}
\hline
\textbf{ADT} & \textbf{User ratio} & \multicolumn{1}{c|}{\textbf{Method}} & \textbf{Recall50} & \textbf{Recall100} & \textbf{NDCG50} & \textbf{NDCG100} \\ \hline \hline
\multirow{6}{*}{T-CE} & \multirow{2}{*}{Ratio < 0.1} & GMF+Warm-up training & 0.0720 & 0.0805 & 0.0201 & 0.0217 \\ 
 &  & GMF+Colliding inference & \textbf{0.0763} & \textbf{0.0890} & \textbf{0.0257} & \textbf{0.0280} \\ \cline{2-7} 
 & \multirow{2}{*}{Ratio < 0.2} & GMF+Warm-up training & \textbf{0.0650} & 0.0924 & 0.0204 & 0.0250 \\ 
 &  & GMF+Colliding inference & 0.0626 & \textbf{0.0959} & \textbf{0.0211} & \textbf{0.0267} \\ \cline{2-7} 
 & \multirow{2}{*}{Ratio < 0.3} & GMF+Warm-up training & 0.0673 & 0.0991 & 0.0231 & 0.0287 \\ 
 &  & GMF+Colliding inference &  \textbf{0.0674} &  \textbf{0.0993} &  \textbf{0.0232} &  \textbf{0.0288} \\ \hline
\multirow{6}{*}{R-CE} & \multirow{2}{*}{Ratio < 0.1} & GMF+Warm-up training & 0.0466 & 0.0805 & 0.0109 & 0.0166 \\
 &  & GMF+Colliding inference & \textbf{0.0678} & \textbf{0.0847} & \textbf{0.0165} & \textbf{0.0195} \\ \cline{2-7} 
 & \multirow{2}{*}{Ratio < 0.2} & GMF+Warm-up training & 0.0571 & 0.0817 & 0.0183 & 0.0224 \\ 
 &  & GMF+Colliding inference & \textbf{0.0625} & \textbf{0.0819} & \textbf{0.0201} & \textbf{0.0234} \\ \cline{2-7} 
 & \multirow{2}{*}{Ratio < 0.3} & GMF+Warm-up training & 0.0619 & 0.0903 & 0.0207 & 0.0257 \\ 
 &  & GMF+Colliding inference & \textbf{0.0625} & \textbf{0.0939} & \textbf{0.0212} & \textbf{0.0267} \\ \hline
\end{tabular}
}
\end{table*}

\vspace{2pt}
\noindent\textbf{Performance Comparison.}
To avoid the detrimental effect of false-positive interactions, a popular idea is to incorporate the extra user feedback (\eg ratings and favorite) for training although they are usually sparse. Existing work either adopts the additional feedback by multi-task learning \cite{Gao2019LearningTR}, or leverages it to identify the true-positive interactions \cite{Wen2019Leveraging, Lu2018Between}. In this work, we introduced two classical models for comparison: Neural Multi-Task Recommendation (NMTR)~\cite{Gao2019LearningTR} and Negative feedback Re-weighting (NR)~\cite{Wen2019Leveraging}. In particular, NMTR considers both click and extra feedback with multi-task learning. NR uses extra feedback (\ie dwell time or rating) to identify true-positive interactions and re-weights the false-positive and non-interacted ones. We applied NMTR, NR, the proposed finetuning, and warm-up training to the testing recommender models and reported the results of GMF in Table \ref{tab:adt_extra}. Other results are omitted due to the similar trend. 
From Table \ref{tab:adt_extra}, we could have the following observations:
\begin{itemize}[leftmargin=*]
    \item NMTR and NR achieve slightly better performance than GMF, which validates the effectiveness of using extra feedback. However, the results of NMTR and NR are inferior to those of the T-CE loss and R-CE loss, which might be attributed to the sparsity of extra feedback. Indeed, the clicks with satisfaction (\ie high ratings) is much fewer than the total number of true-positive interactions. Hence, NR will lose extensive positive training interactions while NMTR will inevitably use many false-positive interactions for training.
    
    \item Directly using extra feedback for finetuning or warm-up training outperforms the vanilla GMF but performs worse than NMTR or NR, showing the advantages of multi-task training~\cite{Gao2019LearningTR} and the re-weighting technique~\cite{Wen2019Leveraging}. 
    
    \item Finetuning and warm-up training significantly improve the performance of ADT over both R-CE and T-CE, further justifying the effectiveness of using extra feedback on denoising methods. Besides, warm-up training is more effective than finetuning. It validates the idea that warm-up training will expose the recommender model to true-positive interactions in the early training stage, and thus improve the model's identification ability during ADT. 
\end{itemize}

\vspace{5pt}
\noindent\textbf{Effectiveness of Colliding Inference.}
Since colliding inference is designed for the users with sparse extra feedback, we leveraged the interaction numbers of extra feedback and implicit feedback to measure the sparsity and divided users into groups. Specifically, we defined \textit{user ratio} as $\frac{\text{the number of extra feedback}}{\text{the number of implicit feedback}}$, and then selected the users with the ratios smaller than the threshold to conduct colliding inference. We chose warm-up training as an example to test colliding inference due to its superior performance, and the results of T-CE and R-CE equipped with warm-up training and colliding inference are summarized in Table \ref{tab:colliding_inference}. 

From Table \ref{tab:colliding_inference}, we could observe that: 
1) colliding inference significantly improves the performance of T-CE and R-CE with warm-up training over the users with low ratios. It validates that distilling colliding effect does utilize extra feedback to extract some useful knowledge from dense users, leading to the better rankings for sparse users; 
and 2) the performance margin is becoming smaller over the user groups from ratio < 0.1 to ratio < 0.3, implying that colliding inference is more effective for sparse users. For the users with high ratios, a larger proportion of extra feedback will prevent the casual effect of $E^*$ on $\hat{E}$ (in Figure \ref{fig:colliding}) from being blocked during ADT.

\begin{figure}[t]
\setlength{\abovecaptionskip}{0cm}
  \centering 
  \hspace{-0.6in}
  \subfigure{
    \includegraphics[width=1.8in]{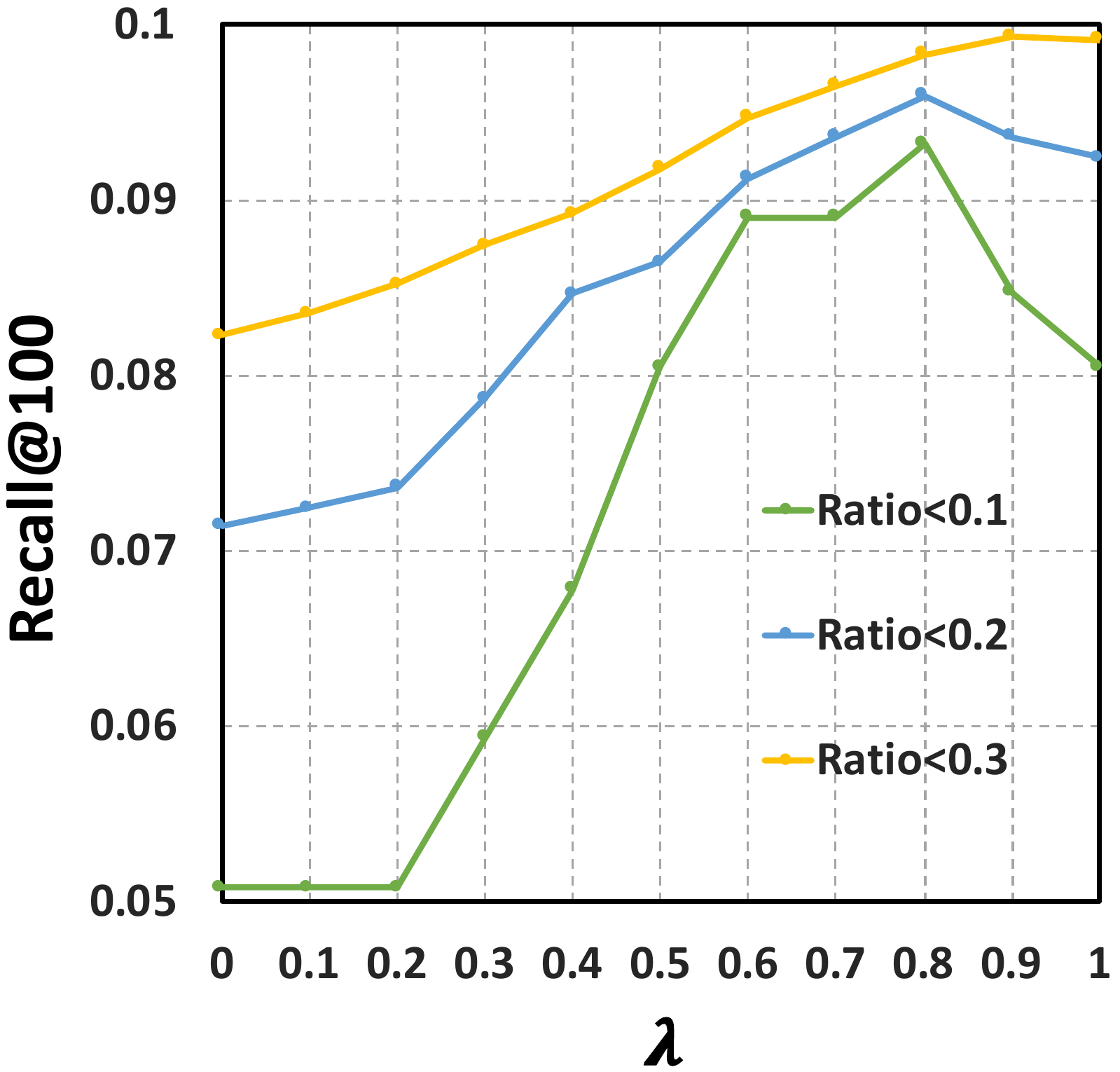}} 
  \hspace{-0.1in}
  \subfigure{
    \includegraphics[width=1.8in]{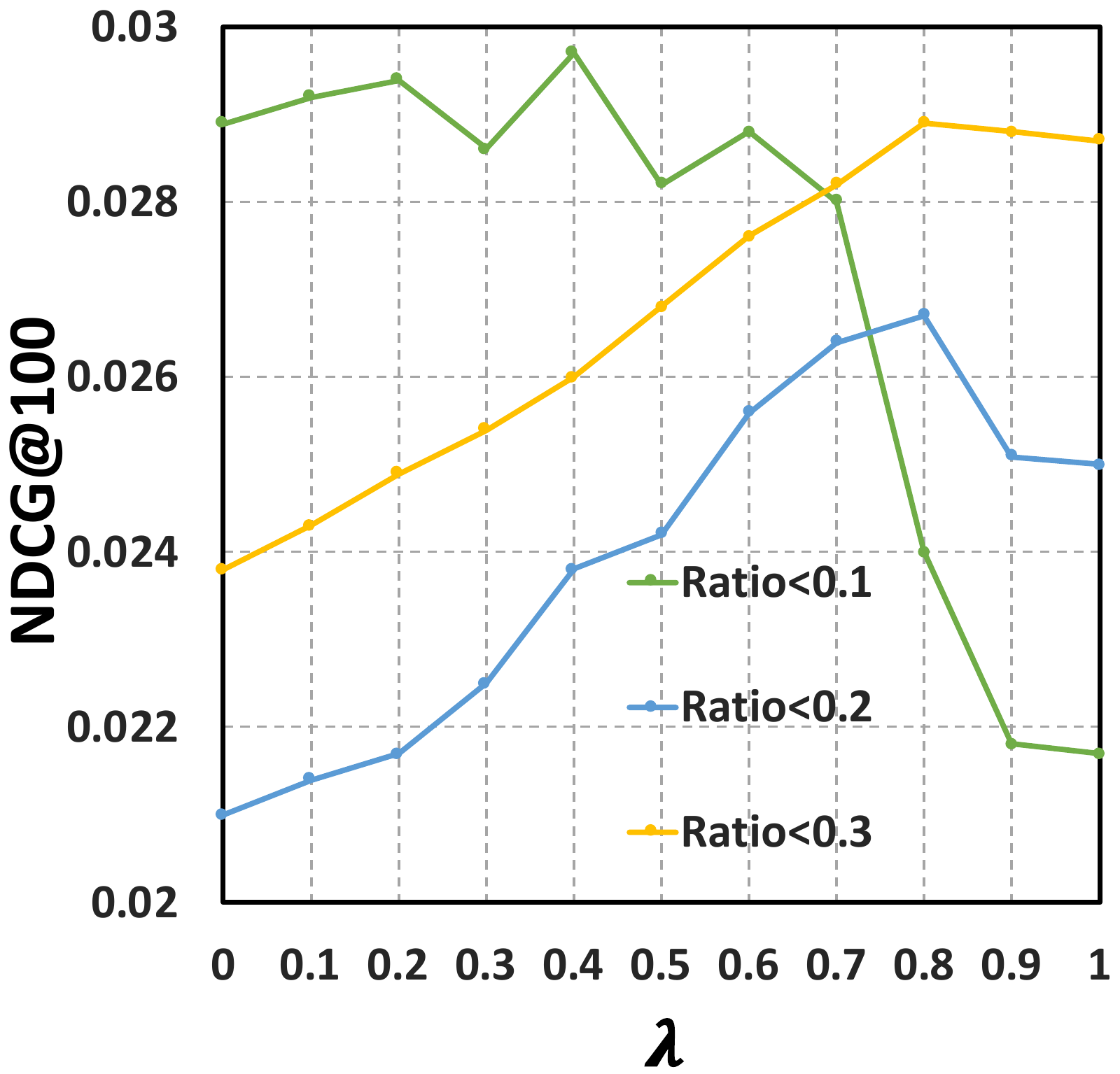}}
  \hspace{-0.6in} 
  \caption{Recall@100 and NDCG@100 of GMF+T-CE with colliding inference \wrt the hyper-parameter $\lambda$.} 
  \label{fig:lambda}
  \vspace{-0.3cm}
\end{figure}

\vspace{5pt}
\noindent\textbf{Hyper-parameter Analysis.}
To study the effect of $\lambda$ in Equation (\ref{eqn:colliding_fusion}), we show the performance comparison over GMF+T-CE \wrt different $\lambda$ values in Figure \ref{fig:lambda}. The results of R-CE are omitted due to the similar trend. From the figure, we could see that: 1) utilizing the prediction scores of neighbors is effective because the best performance happens when $\lambda < 1$. It further validates the effectiveness of colliding inference. And 2) the peaks of the curves occur at $\lambda > 0.5$ in most cases, and move close to 1 as the ratio increases. This first shows that the personalization is essential although the neighbors are selected from the reliable user representation space. Besides, distilling the knowledge from neighbors is more important for the users with sparse extra feedback (\ie low ratio) because the causal effect of extra feedback on their prediction scores will be easily reduced during ADT with noisy implicit feedback.

\section{Conclusion}\label{sec:conclusion}

In this work, we explored denoising implicit feedback for recommender learning. We found the negative effects of noisy implicit feedback, and proposed adaptive denoising training to reduce their impact. In particular, this work contributes two paradigms to formulate the loss functions: Truncated Loss and Reweighted Loss. 
Both paradigms are general and can be applied to different recommendation loss functions, neural recommender models, and optimizers. In this work, we applied the two paradigms on the widely used binary cross-entropy loss and conducted extensive experiments over three recommender models on three datasets, showing that the paradigms could effectively reduce the disturbance of noisy implicit feedback. In addition, we also developed three strategies: finetuning, warm-up training, and colliding inference, which can incorporate extra feedback to further improve the denoising performance of the two paradigms.

This work formulates the task of denoising implicit feedback for recommendation, which points to some new research directions. Specifically, adaptive denoising training is not specific to the recommendation task, and it can be widely used to denoise implicit interactions in other domains, such as Web search and question answering. 
Besides, it is interesting to explore how the proposed two paradigms perform on other loss functions.
Lastly, we propose a novel causality-based inference strategy to improve the ranking performance over sparse users, which inspires the exploration of causal inference into the inference stage of more tasks.


{
\bibliographystyle{IEEEtran}
\bibliography{bibtex}
}

\end{document}